\documentclass[12pt]{article}
\usepackage{epsf,a4,latexsym}
\usepackage[small,bf]{caption}

\newcommand{\half}{\mbox{\small $\frac{1}{2}$}}          
\newcommand{\quart}{\mbox{\small $\frac{1}{4}$}}         
\newcommand{\third}{\mbox{\small $\frac{1}{3}$}}         
\newcommand{\twelfth}{\mbox{\small $\frac{1}{12}$}}      
\newcommand{\msbar}{\mbox{\tiny $\overline{MS}$}}        
\newcommand{\ripmom}{\mbox{\tiny $R\!I^\prime\!\!-\!\!M\!O\!M$}} 
\newcommand{\rgi}{\mbox{\tiny $R\!G\!I$}}                
\newcommand{\trbpt}{\mbox{\tiny $T\!R\!B\!\!-\!\!P\!T$}} 
\newcommand{\lat}{\mbox{\tiny $L\!A\!T$}}                
\newcommand{\con}{\mbox{\tiny $C\!O\!N$}}                
\newcommand{\qed}{\mbox{\tiny $Q\!E\!D$}}                
\newcommand{\plaq}{\Box}                                 
\newcommand{\born}{\mbox{\tiny $B\!O\!R\!N$}}            
\newcommand{\expt}{\mbox{\tiny $E\!X\!P\!T$}}            
\newcommand{\bare}{\mbox{\tiny $bare$}}                  
\newcommand{\smeared}{\mbox{\tiny $smeared$}}            

\def\lsim{\mathrel{\rlap{\lower4pt\hbox{\hskip1pt$\sim$}}
    \raise1pt\hbox{$<$}}}                
\def\gsim{\mathrel{\rlap{\lower4pt\hbox{\hskip1pt$\sim$}}
    \raise1pt\hbox{$>$}}}                
\begin{document}

\title{
\vspace{-2.5cm}
\flushleft{\normalsize DESY 05-237} \\
\vspace{-0.35cm}
{\normalsize Edinburgh 2005/18} \\
\vspace{-0.35cm}
{\normalsize Liverpool LTH 672} \\
\vspace{-0.35cm}
{\normalsize January 2006} \\
\vspace{0.5cm}
\centering{\Large \bf Estimating the Unquenched Strange Quark Mass from
                      the Lattice Axial Ward Identity}}

\author{\large M. G\"ockeler$^1$, R. Horsley$^2$, A.~C. Irving$^3$,  \\
               D. Pleiter$^4$, P.~E.~L. Rakow$^3$, G. Schierholz$^{4,5}$, \\
               H. St\"uben$^6$  and J.~M. Zanotti$^2$ \\[1em]
         -- QCDSF-UKQCD Collaboration -- \\[1em]
        \small
           $^1$ Institut f\"ur Theoretische Physik,
                Universit\"at Regensburg, \\[-0.5em]
        \small
                D-93040 Regensburg, Germany \\[0.25em]
        \small
           $^2$ School of Physics, University of Edinburgh, \\[-0.5em]
        \small 
                Edinburgh EH9 3JZ, UK \\[0.25em]
        \small
           $^3$ Department of Mathematical Sciences, University of Liverpool,
                                                            \\[-0.5em]
        \small
                Liverpool L69 3BX, UK \\[0.25em]
        \small
           $^4$ John von Neumann-Institut f\"ur Computing NIC, \\[-0.5em]
        \small
                Deutsches Elektronen-Synchrotron DESY, \\[-0.5em]
        \small
                D-15738 Zeuthen, Germany \\[0.25em]
        \small
           $^5$ Deutsches Elektronen-Synchrotron DESY, \\[-0.5em] 
        \small
                D-22603 Hamburg, Germany \\[0.25em]
        \small
           $^6$ Konrad-Zuse-Zentrum f\"ur Informationstechnik Berlin,
                                                       \\[-0.5em]
        \small
                D-14195 Berlin, Germany}

\date{}

\maketitle


\begin{abstract}
We present a determination of the strange quark mass for two flavours
($n_f=2$) of light dynamical quarks using the axial Ward identity.
The calculations are performed on the lattice using $O(a)$
improved Wilson fermions and include a fully non-perturbative determination
of the renormalisation constant.
In the continuum limit we find
$m_s^{\msbar}(2\,\mbox{GeV}) = 111(6)(4)(6) \, \mbox{MeV}$, using the
force scale $r_0 = 0.467\,\mbox{fm}$, where the first error is statistical,
the second and third are systematic due to the fit and scale uncertainties
respectively. Results are also presented for the light quark mass
and the chiral condensate. The corresponding results are also given for
$r_0=0.5\,\mbox{fm}$.

\end{abstract}

\clearpage


\section{Introduction}
\label{introduction}


Lattice methods allow, in principle, an `ab initio'
calculation of the fundamental parameters of QCD, among them
the quark masses. Quarks are not asymptotic states of QCD and
so quark masses need to be defined by a renormalisation procedure,
\begin{equation}
   m_q^{\cal S}(M) = Z_m^{\cal S}(M) m_q^{\bare} \,,
\end{equation}
by giving the scheme ${\cal S}$ and scale $M$.

To convert the lattice results to continuum numbers, one needs control
over the discretisation errors and the matching relations between
the lattice scheme and the continuum renormalisation scheme $\cal S$.
Discretisation errors can be kept small and manageable by employing
an improved fermion action. But, still, the lattice numbers may show
considerable cut-off dependence at present couplings, which requires
that the calculations are done over a range of sufficiently small
lattice spacings $a$, discretisation errors then being removed
by an extrapolation to $a=0$. The perturbative relations
between renormalised quantities in the continuum and the bare lattice
results are in almost all cases known to one-loop order only. Data show
(for example \cite{bakeyev03a}) that $O(\alpha_s^2)$ corrections
can be large, $O(10-20\%)$ at spacings $a \approx 0.1\,\mbox{fm}$,
which makes a non-perturbative calculation of the renormalisation
constants indispensible.

Many calculations of the strange quark mass, both with $n_f=2$
\cite{alikhan00a,alikhan01a,aoki02a} and $n_f=2+1$
\cite{kaneko03a,aubin04a,ishikawa05a,mason05a} flavours of sea quarks,
employed perturbative renormalisation of the bare quark mass and were
restricted to lattice spacings $a \gsim 0.1\,\mbox{fm}$.
(Recent $n_f=2+1$ results \cite{ishikawa05a}, have finer lattice
spacings.) These authors quote strange quark
masses of $O(80)\,\mbox{MeV}$, lying $15-30\%$ below the corresponding
quenched results \cite{gockeler99a,garden99a}.

In \cite{gockeler04a} we have presented an entirely non-perturbative (NP)
calculation of the light quark masses based on the vector Ward identity (VWI),
using non-perturbatively $O(a)$ improved Wilson fermions with $n_f=2$
flavours of dynamical quarks. The calculation was done at four different
lattice spacings $0.065 \lsim a \lsim 0.09\, \mbox{fm}$, which
allowed us to perform a continuum extrapolation. We found a strange
quark mass of $m_s^{\msbar}(2\, \mbox{GeV}) = 119(5)(8)$ MeV.
A highlight and essential ingredient of the calculation was that
we were able to compute the flavour singlet mass renormalisation constant,
which is needed in the VWI approach.

This result has been complemented recently by further studies in
\cite{dellamorte05a,becirevic05a} which also used NP determinations
of the renormalisation constant.

In this paper we present an independent calculation of the strange quark mass
using the axial vector Ward identity (AWI), again for $n_f=2$ flavours of
improved Wilson fermions. The AWI involves only non-singlet quantities and
thus provides an important test of our previous calculation.

The paper is organised as follows. As we shall be considering not only
the $\overline{\rm MS}$ scheme, but also the $\rm{RI}^\prime$-MOM scheme
(which is more convenient for a lattice calculation) we first discuss
in section~\ref{rgi} renormalisation group invariants (RGIs),
taking the quark mass as an example, and how to convert to them.
We also collect together relevant formulae for the $\overline{\rm MS}$
scheme. Also discussed is the unit and scale we shall
use -- the $r_0$-force scale -- and thus the relevant conversion
factor to physical units. In section~\ref{cpt} we compile some results
from leading order (LO) and next to leading order (NLO)
chiral perturbation theory ($\chi$PT) and re-write them in a form
suitable for our calculation. Section~\ref{lattice} describes some
lattice details relevant for $O(a)$ improved fermions.
This is followed in section~\ref{renormalisation}
by the non-perturbative computation of the renormalisation constant.
Relevant results for the $\rm{RI}^\prime$-MOM scheme are given,
both for the the lattice computation of $Z_m^{\ripmom}$ and for
the conversion to the RGI form. The section is concluded with a
comparison of this result with the results obtained by other
approaches (principally the tadpole improved (TI) perturbation
theory method).
In section~\ref{results} results are given for the strange quark mass,
first at finite lattice spacing, and then the continuum extrapolation
is performed to give our final answer.
Finally in the last section, section~\ref{conclusions},
we compare our AWI result with the previously obtained VWI result 
and also with other recent mass determinations. In the appendix,
tables of our raw data results for the quark mass are given.


\section{Renormalisation Group Invariants}
\label{rgi}


The `running' of the renormalised quark mass as the scale $M$
is changed is controlled by the $\beta$ and $\gamma$ functions
in the renormalisation group equation, defined by
\begin{eqnarray}
   \beta^{\cal S} \left(g^{\cal S}(M) \right) &\equiv&
                 \left. {\partial g^{\cal S}(M) \over
                         \partial \log M }\right|_{\bare},
                                   \label{beta_def} \\
   \gamma_m^{\cal S} \left(g^{\cal S}(M) \right) &\equiv&
                 \left. {\partial \log Z_m^{\cal S}(M) \over
                         \partial \log M }\right|_{\bare},
                                   \label{gamma_def}
\end{eqnarray}
where the bare parameters are held constant. These functions are given
perturbatively as power series expansions in the coupling constant,
\begin{eqnarray}
   \beta^{\cal S}(g)    &=& - b_0g^3 - b_1g^5
                            - b_2^{\cal S}g^7 - b_3^{\cal S}g^9 - \ldots \,,
                                    \nonumber \\
   \gamma_m^{\cal S}(g) &=&   d_{m0}g^2 + d_{m1}^{\cal S}g^4
                            + d_{m2}^{\cal S}g^6 + d_{m3}^{\cal S}g^8 +
                                          \ldots \,.
\end{eqnarray}
The first two coefficients of the $\beta$-function and first coefficient
of the $\gamma_m$ function are scheme independent,
\begin{equation}
   b_0 = {1\over (4\pi)^2}
           \left( 11 - {2\over 3}n_f \right) \,, \qquad
   b_1 = {1\over (4\pi)^4}
           \left( 102 - {38 \over 3} n_f \right) \,.
\label{b0+b1}
\end{equation}
and
\begin{equation} 
   d_{m0} = - { 8 \over (4\pi)^2} \,,
\end{equation}
while all others depend on the scheme chosen.

We may immediately integrate eq.~(\ref{beta_def}) to obtain
\begin{equation}
   { M \over\Lambda^{\cal S} }
      =  \left[b_0 g^{\cal S}(M)^2
                          \right]^{b_1\over 2b_0^2}
         \exp{\left[{1\over 2b_0 g^{\cal S}(M)^2}\right]} 
         \exp{\left\{ \int_0^{g^{\cal S}(M)} d\xi
          \left[ {1 \over \beta^{\cal S}(\xi)} +
                 {1\over b_0 \xi^3} - {b_1\over b_0^2\xi}
         \right]\right\} } \,.
\label{lambda_def}
\end{equation}
The renormalisation group invariant quark mass%
\footnote{Analogous definitions hold for other quantities which depend
on the scheme and scale chosen.}
is defined from the renormalised quark mass as
\begin{equation}
   m_q^{\rgi} \equiv \Delta Z_m^{\cal S}(M) m^{\cal S}(M)
               = \Delta Z_m^{\cal S}(M) Z_m^{\cal S}(M) m_q^{\bare}
               \equiv Z_m^{\rgi} m_q^{\bare} \,,
\label{mrgi_msbar}
\end{equation}
where
\begin{equation}
   [\Delta Z_m^{\cal S}(M)]^{-1} = 
          \left[ 2b_0 g^{\cal S}(M)^2 \right]^{-{d_{m0}\over 2b_0}}
          \exp{\left\{ \int_0^{g^{\cal S}(M)} d\xi
          \left[ {\gamma_m^{\cal S}(\xi)
                             \over \beta^{\cal S}(\xi)} +
                 {d_{m0}\over b_0 \xi} \right] \right\} },
\label{deltam_def}
\end{equation}
and so the integration constant upon integrating eq.~(\ref{beta_def})
is given by $\Lambda^{\cal S}$, and similarly from eq.~(\ref{gamma_def})
the integration constant is $m_q^{\rgi}$. $\Lambda^{\cal S}$ and
$m_q^{\rgi}$ are thus independent of the scale. (Note that although
the functional form of $\Delta Z_m^{\cal S}(M)$ is fixed,
the absolute value is not; conventions vary for its definition.)
Also for a scheme change ${\cal S}\to {\cal S}^\prime$
(it is now sufficient to take them at the same scale) given by
\begin{equation}
   g^{{\cal S}^\prime} = G(g^{\cal S}) = 
             g^{\cal S}(1 + \half t_1 (g^{\cal S})^2 + \ldots) \,,
\label{G_def}
\end{equation}
then $m_q^{\rgi}$ remains invariant, while $\Lambda$ changes as
$\Lambda^{{\cal S}^\prime} = \Lambda^{\cal S} \exp(t_1/(2b_0))$.
Note also that analytic expressions for the integrals in 
eq.~(\ref{mrgi_msbar}) or eq.~(\ref{deltam_def}) can be found for
low orders, for example to two loops we have
\begin{equation}
   \Delta Z^{\cal S}_m(M) =
      \left[ 2 b_0 (g^{\cal S}(M))^2 \right]^{d_{m0}\over 2b_0}
      \left[ 1 + {b_1 \over b_0} (g^{\cal S}(M))^2
      \right]^{{b_0 d_{m1}^{\cal S} - b_1 d_{m0}
                                               \over 2 b_0 b_1} } \,.
\label{DelZ_2loop}
\end{equation}

Thus we have a convenient splitting of the problem into two parts:
a number, $m_q^{\rgi}$, which involves a non-perturbative computation,
and is the goal of this paper and, if desired, an evaluation of
$\Delta Z_m^{\cal S}$ which allows the running quark mass to be given
in a renormalisation scheme ${\cal S}$.

In the remainder of this section we discuss the evaluation of
$\Delta Z_m^{\cal S}$ in the $\overline{\rm MS}$-scheme, which is
conventionally used, and for which four coefficients in the
perturbative expansion are known. For the $\beta$ function we have
\cite{tarasov80a,larin93a,vanritbergen97a},
\begin{eqnarray}
   b_2^{\msbar}
        &=& { 1\over (4\pi)^6}\left[ {2857 \over 2} - 
                                     {5033 \over 18}n_f +
                                     {325 \over 54}n_f^2
                              \right]     \,,
                                    \nonumber \\
   b_3^{\msbar}
        &=& {1 \over (4\pi)^8}\left[ {149753 \over 6} + 3564\zeta_3 -
                                     ( {1078361 \over 162} +
                                       {6508 \over 27}\zeta_3 )n_f \right.
                                    \nonumber \\
        & &  \hspace*{1.75in} \left. + ( {50065 \over 162} + {6472 \over 81}
                                           \zeta_3 ) n_f^2 +
                                     {1093 \over 729} n_f^3 
                              \right] \,,
\label{b2+3_msbar}
\end{eqnarray}
and for the $\gamma_m$ function \cite{chetyrkin97a,vermaseren97a},
\begin{eqnarray}
   d_{m1}^{\msbar}
        &=& - { 1 \over (4\pi)^4}\left[
                     {404 \over 3} - {40 \over 9}n_f
                                 \right] \,,
                                              \\
   d_{m2}^{\msbar} 
        &=& - { 1 \over (4\pi)^6}\left[
                    2498 - ({4432 \over 27} + {320 \over 3}\zeta_3) n_f
                          - {280 \over 81}n_f^2 
                                 \right] \,,
                                    \nonumber \\
   d_{m3}^{\msbar}
        &=& - { 1 \over (4\pi)^8}\left[
                     {4603055 \over 81} + {271360 \over 27}\zeta_3 
                                        - 17600\zeta_5  \right.
                                    \nonumber \\
        & &  \hspace*{1.00in}    \left.
                   -( {183446 \over 27} + {68384 \over 9}\zeta_3
                     - 1760\zeta_4 - {36800 \over 9}\zeta_5 ) n_f \right.
                                    \nonumber \\
        & &  \hspace*{1.00in}    \left.
                   +( {10484 \over 243} + {1600 \over 9}\zeta_3 
                      - {320 \over 3}\zeta_4 ) n_f^2
                   -(  {664 \over 243} - {128 \over 27}\zeta_3 ) n_f^3 )
                                 \right] \,,
                                    \nonumber
\label{d1+2+3_msbar}
\end{eqnarray}
with $\zeta_3=1.20206\ldots$, $\zeta_4=1.08232\ldots$  and
$\zeta_5= 1.03693\ldots$, $\zeta$ being the Riemann zeta function.

\begin{figure}[t]
   \hspace{0.50in}
      \epsfxsize=10.00cm \epsfbox{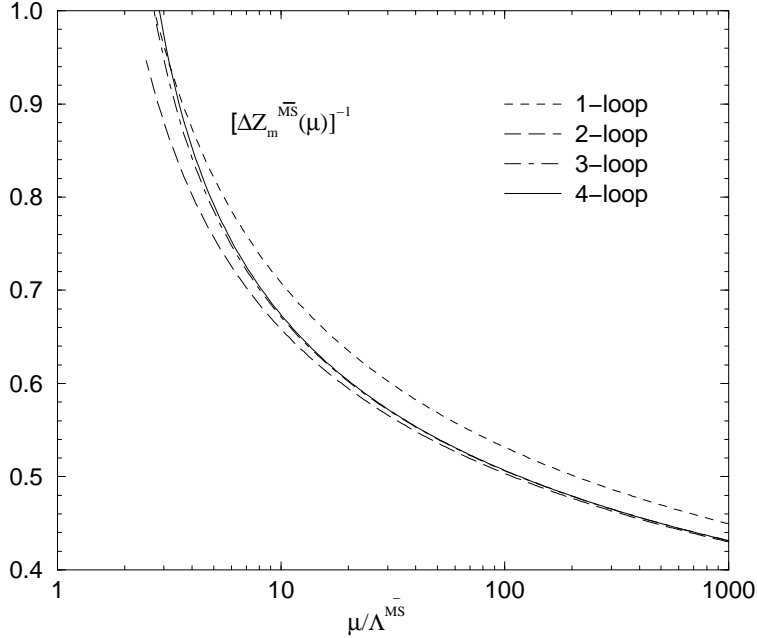}
   \caption{One-, two-, three- and four-loop results for
            $[\Delta Z_m^{\msbar}(\mu)]^{-1}$ in units of
            $\Lambda^{\msbar}$.}
   \label{fig_Del_Zm_MSbar_nf2_muolam}
\end{figure}

This scheme is a manifestly perturbative scheme
and so should be used at a high enough scale $M  \equiv \mu$
so that perturbation theory is reliable.
Computing $[\Delta Z_m^{\msbar}(\mu)]^{-1}$ involves first solving
eq.~(\ref{lambda_def}) for $g^{\msbar}$ (as a function of
$\mu / \Lambda^{\msbar}$) and then evaluating eq.~(\ref{deltam_def}).
Practically we expand the $\beta$ and $\gamma$ functions
to the appropriate order and then numerically evaluate the integrals.
The final results are given in Fig.~\ref{fig_Del_Zm_MSbar_nf2_muolam}.

Conventionally light quark masses are defined at a scale of
$\mu = 2\,\mbox{GeV}$, which means giving a value for $\Lambda^{\msbar}$
in $\mbox{MeV}$. We set the scale here by using the `force scale'
$r_0$, which means first changing from the $\Lambda^{\msbar}$ unit
to the $r_0$ unit. From \cite{gockeler05a} (see also \cite{dellamorte04a}),
we use the value
\begin{equation}
   r_0 \Lambda^{\msbar} = 0.617(40)(21) \,.
\label{r0lam_value}
\end{equation}
The $r_0$ scale in $\mbox{MeV}$ still needs to be set. A popular choice is
\begin{equation}
   r_0 = 0.5 \,\mbox{fm} \equiv 1/(394.6\,\makebox{MeV}) \,,
\label{r0_0.500_phys}
\end{equation}
which is useful when making comparisons with other results
for the quark mass. Alternatively from a fit to the dimensionless
nucleon mass $r_0 m_N$ using results obtained by the CP-PACS, JLQCD
and QCDSF-UKQCD collaborations, following \cite{alikhan03a}
we found a scale of
\begin{equation}
   r_0 = 0.467(33) \,\mbox{fm} \equiv 1/(422.5(29.9)\,\makebox{MeV}) \,.
\label{r0_0.467_phys}
\end{equation}
Similar results were obtained in \cite{aubin04b,gockeler05c}.
The $0.033\, \mbox{fm}$ error in the $r_0 = 0.467 \,\mbox{fm}$ estimate
is roughly equal to the difference between the two $r_0$ values.
As these different $r_0$ values give an idea of the uncertainties involved
in setting the scale we shall derive results using both values of $r_0$
and regard this as giving an estimate of a possible scale systematic error.
The different values of $r_0$ in eqs.~(\ref{r0_0.500_phys}),
(\ref{r0_0.467_phys}) give $\Lambda^{\msbar} = 243(16)(8) \, \mbox{MeV}$
and $261(17)(9) \, \mbox{MeV}$ respectively.

Results for $[\Delta Z_m^{\msbar}]^{-1}$ at $\mu = 2\,\mbox{GeV}$
are given in Table~\ref{table_2GeVmsbar_values}.
\begin{table}[t]
   \begin{center}
      \begin{tabular}{||l||l|l|l|l||}
         \hline
         \multicolumn{1}{||c||}{$r_0$}
                           & one-loop & two-loop & three-loop & four-loop  \\
         \hline
         \hline
  $0.5      \, \mbox{fm}$  & $0.735(11)$ & $0.682(10)$
                           & $0.698(11)$ & $0.700(11)$  \\
  $0.467    \, \mbox{fm}$  & $0.745(12)$ & $0.690(10)$
                           & $0.707(12)$ & $0.711(12)$  \\
         \hline
         \hline
      \end{tabular}
   \end{center}
\caption{Values of $[\Delta Z_m^{\msbar}(\mu)]^{-1}$ at $\mu = 2\,\mbox{GeV}$.
         The errors are  a reflection of the errors in
         eq.~(\ref{r0lam_value}).}
\label{table_2GeVmsbar_values}
\end{table}
At $\mu = 2\,\mbox{GeV}$ we have $\mu / \Lambda^{\msbar} \approx 8$,
and it seems that already at this value we have a rapidly
converging series in loop orders. Indeed, only going from one loop
to two loops gives a significant change in
$[\Delta Z_m^{\msbar}]^{-1}$ of order $8\%$.
From two loops to three loops we have about $2\%$. The difference
between the three-loop and four-loop results is $O(\half\%)$.
So if we are given $m_q^{\rgi}$, and we wish to find the quark mass in the
$\overline{\rm MS}$ scheme at a certain scale, we shall use the four-loop
result from eq.~(\ref{deltam_def}) as shown in 
Fig.~\ref{fig_Del_Zm_MSbar_nf2_muolam}.


\section{Chiral Perturbation Theory}
\label{cpt}


Chiral perturbation theory ($\chi$PT) gives the LO and NLO result%
\footnote{The NNLO result has recently been constructed in
\cite{bijnens05a}.}
\cite{bernard93a,sharpe97a} for $n_f$ degenerate sea quarks of
\begin{eqnarray}
   \left( { m^{AB}_{ps} \over 4\pi f_0 } \right)^2 
      &=& \chi_{AB} \left[ 1 + \chi_S n_f(2\alpha_6 - \alpha_4) 
                        + \chi_{AB}(2\alpha_8 - \alpha_5) 
                                \phantom{1 \over n_f}\right.
                                                 \nonumber    \\
      & & \hspace*{0.50in} \left.
               + {1 \over n_f} { \chi_A(\chi_S - \chi_A)\ln \chi_A -
                                 \chi_B(\chi_S - \chi_B)\ln \chi_B
                                          \over  \chi_B - \chi_A }
               \right] \,,
\label{NLO_cpt}
\end{eqnarray}
where the first term on the RHS is the LO term, the NLO terms being
the remaining terms and $\alpha_i$ are the low energy chiral constants (LECs)
evaluated at the scale $\Lambda_{\chi} = 4\pi f_\pi$. With our
conventions for the pion decay constant in the chiral limit,
$f_0$, we have $f_\pi = 92.4\,\mbox{MeV}$ so that
$\Lambda_{\chi} \approx 1160\,\mbox{MeV}$.

$\chi_{AB}$ are related to the quark masses by \cite{sharpe97a},
\begin{equation}
   \chi_{AB} = { B_0^{\cal S} ( m_A + m_B )^{\cal S} \over
              (4\pi f_0 )^2 }\,, \qquad A, B \in \{ V_1, V_2, S \} \,,
\label{chi_def}
\end{equation}
where $m_S$ is the sea quark mass and $m_{V_i}$, $i = 1, 2$ are the
(possibly) non-degenerate valence quark masses. In particular we have
\begin{equation}
   \chi_{AB} = \half( \chi_A + \chi_B ) \,, 
               \qquad \mbox{where} \qquad \chi_A \equiv \chi_{AA}  \,.
\label{chi_AA_def}
\end{equation}
Furthermore in eq.~(\ref{chi_def}), $B_0^{\cal S}$ is related to the
chiral condensate by
\begin{equation}
   B_0^{\cal S} = - {1 \over f_0^2} \langle \overline{q}q \rangle^{\cal S} \,.
\label{chiral_condensate}
\end{equation}
Apart from $f_\pi$, none of the other LECs (here $B_0$, or
$\langle \overline{q}q \rangle$, and $\alpha_4, \alpha_5, \alpha_6, \alpha_8$
in eq.~(\ref{NLO_cpt})) are well determined. Typical values are (for $n_f=3$), 
$\alpha_4 = - 0.76(60)$, $\alpha_5 = 0.5(6)$, $\alpha_6 = - 0.5(4)$,
$\alpha_8 = 0.76(40)$ (as compiled in \cite{heitger00a})
giving
\begin{equation}
   2\alpha_6 - \alpha_4 \approx -0.24 \,, \quad
   2\alpha_8 - \alpha_5 \approx -1.02 \,, \quad
   \langle \overline{q}q \rangle^{\msbar}(2\,\mbox{GeV})
                        \approx -(267\,\mbox{MeV})^3 \,,
\label{LEC_alpha}
\end{equation}
(the $\langle \overline{q}q \rangle$ result is taken from \cite{jamin02a}).

$m^{AB}_{ps}$ in eq.~(\ref{NLO_cpt}) is the pseudoscalar mass
(with $A, B \in \{ V_1, V_2 \}$) implicitly depending on the
sea quark mass $m_S$. Again for degenerate valence quarks, we write 
$m^A_{ps} \equiv m^{AA}_{ps}$. Note that we also numerically allow
for a valence quark mass to be equal to the sea quark mass, so 
for example we can write $m_{ps}^S$.

We shall assume eq.~(\ref{NLO_cpt}) as the basic functional form for
the relation between the quark mass and the pseudoscalar mass in the
following. As expected this equation is symmetric under an interchange
of the two valence quarks. However it does not have the most general
structure allowed by this symmetry.

Eq.~(\ref{NLO_cpt}) is not very convenient for comparing with 
numerical results for a variety of reasons
(see section~\ref{generalities} and eqs.~(\ref{y+M_def}),
(\ref{fit_function})).
First, the quantities we measure are not $\chi_{AB}$, $m^{AB}_{ps}$
but only proportional to them,
\begin{equation}
   \chi_{AB} = c_{\chi} y_{AB} \,, \qquad 
   {m^{AB}_{ps} \over 4\pi f_0} = c_m M^{AB}_{ps} \,,
\label{y_M_defs}
\end{equation}
where $y_{AB}$ and $M^{AB}_{ps}$ are new variables.
Substituting these into eq.~(\ref{NLO_cpt}) simply shifts
the coefficients of the various terms (including the $1$ and $1/n_f$
terms), while the structural form of this equation remains the same.
Note that this includes the cases where $c_{\chi}$ and $c_m$
are functions of the quark mass, e.g.
\begin{eqnarray}
   c_{\chi} &\to&
     c_{\chi}^{(0)} + c_{\chi}^{S(1)}y_S + c_{\chi}^{AB(1)}y_{AB} + \ldots \,,
                                           \nonumber \\
   c_m      &\to&
     c_m^{(0)} + c_m^{S(1)}y_S + c_m^{AB(1)}y_{AB} + \ldots \,.
\label{c_general}
\end{eqnarray}
Secondly, we prefer to work with a function of the pseudoscalar mass
rather than the quark mass, so we invert eq.~(\ref{NLO_cpt}).
This gives a functional form, which we choose to write as
\begin{eqnarray}
   { y_{AB} \over (M^{AB}_{ps})^2 } 
      &=& c_a  + \left( {c_b - c_d (1+\ln c_a)\over c_a} \right) y_{S}
               + \left( {c_c + c_d(1+2\ln c_a) \over c_a} \right) y_{AB} 
                                                     \nonumber \\
      & & \hspace{0.5in}
             - \left({c_d \over c_a} \right) { y_A(y_S - y_A)\ln y_A -
                                 y_B(y_S - y_B)\ln y_B
                                          \over y_B - y_A } \,.
\label{non-degenerate_cPT}
\end{eqnarray}
Setting $A = B = V$ this equation reduces to the degenerate valence
case (and finally setting $V \equiv S$ gives the sea quark case).
These sets of equations may be (once) iterated to produce 
$y_{AB}/(M^{AB}_{ps})^2$ as a function of $(M^{AB}_{ps})^2$ and
$(M^S_{ps})^2$. Relevant later will be the case of degenerate
valence quarks when we have
\begin{equation}
   { y_V \over (M^V_{ps})^2 } 
      = c_a + c_b (M^S_{ps})^2 + c_c (M^V_{ps})^2
            + c_d \left( (M^S_{ps})^2 - 2(M^V_{ps})^2 \right)
                                  \ln (M^V_{ps})^2 \,,
\label{cpt_VV}
\end{equation}
which explains our original choice of the $c_a$, $c_i$ ($i = b, c, d$)
coefficients in eq.~(\ref{non-degenerate_cPT}). So, as mentioned previously,
we see that determining these coefficients from eq.~(\ref{cpt_VV})
which only needs degenerate valence quark masses is sufficient
to find the results for non-degenerate quark masses,
eq.~(\ref{non-degenerate_cPT}).

Numerically we shall find that higher order terms in $\chi$PT are small,
i.e.\ $c_a \gg |c_i|M^2_{ps}$ ($i = b, c, d$) and thus all these
manipulations are justified. 

The relation between $c_\chi$, $c_m$ and $c_a$ is, from the LO term
\begin{equation}
   {c_m^2 \over c_\chi} = c_a \,.
\label{LEC_LO_relation}
\end{equation}
At NLO we have in addition%
\begin{equation}
   c_m^2 = n_f { c_d \over c_a} \,,
\label{LEC_NLO_cm_relation}
\end{equation}
and for the remaining LECs the relationships
\begin{eqnarray}
   2\alpha_6 - \alpha_4
         &=&   {1\over n_f^2} \left[
               1 + \ln n_f - {c_b \over c_d} + \ln {c_d \over c_a}
                           \right] \,,
                                                     \nonumber \\
   2\alpha_8 - \alpha_5
        &=& - {1\over n_f} \left[
              1 + 2\ln n_f + {c_c \over c_d} + 2\ln {c_d \over c_a}
                           \right] \,.
\label{LEC_c_relation}
\end{eqnarray}
We write the results only for the case when $c^{(1)}_m = c^{(1)}_{\chi} = 0$
(see eq.~(\ref{c_general})). This is sufficient for the later estimation
of the LECs in  section~\ref{results}, as we shall be using continuum
results%
\footnote{ For the more general case in
eqs.~(\ref{LEC_LO_relation}), (\ref{LEC_NLO_cm_relation}) and
(\ref{LEC_c_relation}) $c_m$, $c_\chi$ are replaced by $c^{(0)}_m$
and $c^{(0)}_\chi$ respectively, together with additional terms on
the RHS of eq.~(\ref{LEC_c_relation}) of
$ - 1/(n_f (c_\chi^{(0)})^2) ( c_\chi^{S(1)} - 2c_m^{S(1)}/c_m^{(0)} )$ and
$ - 1/(c_\chi^{(0)})^2 ( c_\chi^{AB(1)} - 2c_m^{AB(1)}/c_m^{(0)} )$ for
the second and third equations, respectively.}.

Finally we have to find a formula for the strange quark mass.
We first note that we have
\begin{itemize}
   \item Two degenerate sea quarks $m_S \equiv m_{ud} = \half ( m_u + m_d)$
   \item Three possible valence quarks, $m_u$, $m_d$ and $m_s$
         (where `$s$' denotes the strange quark). We write
         $m_u = m_{ud} - \Delta m_{ud}$ and $m_d = m_{ud} + \Delta m_{ud}$
         where $\Delta m_{ud} = (m_d - m_u)/2$ is proportional to the
         difference between the down and up quark masses.
\end{itemize}
Inputting meson data we use the $K^+$ ($u\bar{s}$), where we set $A=u$,
$B=s$ in eq.~(\ref{non-degenerate_cPT}), $K^0$ ($d\bar{s}$),
where we have $A=d$, $B=s$, and together with the $\pi^+$ ($u\bar{d}$),
with $A=u$, $B=d$, gives after some algebra the result
\begin{eqnarray}
   y_s &=& c_a \left[ M_{K^+}^2 + M_{K^0}^2 - M_{\pi^+}^2 \right]
                                           \nonumber \\
       & & + (c_b-c_d) \left[M_{K^+}^2 + M_{K^0}^2 \right]M_{\pi^+}^2 
                                           \nonumber \\
       & & + \half (c_c+c_d)\left[M_{K^+}^2 + M_{K^0}^2 \right]^2
                                           \nonumber \\
       & & - (c_b+c_c)M_{\pi^+}^4
                                           \nonumber \\
       & & - c_d \left[M_{K^+}^2 + M_{K^0}^2 \right]
                 \left[ M_{K^+}^2 + M_{K^0}^2 - M_{\pi^+}^2 \right]
                 \ln \left( M_{K^+}^2 + M_{K^0}^2 - M_{\pi^+}^2 \right)
                                           \nonumber \\
       & & + c_d M_{\pi^+}^4 \ln M_{\pi^+}^2  + \cdots \,,
\label{strange_general}
\end{eqnarray}
for the strange quark and
\begin{equation}
   y_{ud} = c_a M_{\pi^+}^2 + (c_b+c_c) M_{\pi^+}^4
                            - c_d M_{\pi^+}^4 \ln M_{\pi^+}^2 + \cdots \,,
\label{ud_general}
\end{equation}
for the light quark, where the $\cdots$ include higher order terms
in $\chi$PT (i.e.\ NNLO) and terms of $O((\Delta m_{ud})^2)$.

The results of eqs.~(\ref{strange_general}) and (\ref{ud_general})
are valid for `pure' QCD. To include electromagnetic effects, we use
Dashen's theorem which says that if electromagetic effects are
the only source of breaking of isospin symmetry (i.e.\ $m_u = m_d$),
the leading electromagnetic contribution to $m_{K^+}^2$ and $m_{\pi^+}^2$
are equal
while $m_{\pi^0}^2$ and $m_{K^0}^2$ are unaffected (see e.g.\ \cite{cheng84a}).
Thus the masses in eqs.~(\ref{strange_general}) and (\ref{ud_general})
may be written as \cite{leutwyler01a}
\begin{eqnarray}
   m_{K^+}^2   &=& (m_{K^+}^{\expt})^2 - (m_{\pi^+}^{\expt})^2
                                       + (m_{\pi^0}^{\expt})^2 \,,
                                           \nonumber \\
   m_{K^0}^2   &=& (m_{K^0}^{\expt})^2  \,,
                                           \nonumber \\
   m_{\pi^+}^2 = m_{\pi^0}^2 &=& (m_{\pi^0}^{\expt})^2 \,.
\label{ps_pureQCD_values}
\end{eqnarray}
where $m_{K^+}$, $m_{K^0}$, $m_{\pi^+}$ and $m_{\pi^0}$ are the
`pure' QCD numbers, while $m_{K^+}^{\expt}$, $m_{K^0}^{\expt}$, 
$m_{\pi^+}^{\expt}$ and $m_{\pi^0}^{\expt}$ are the experimentally
observed numbers. Dashen's theorem has corrections $O(\alpha_{\qed}m_q)$ 
from higher order terms in $\chi$PT, estimates vary as to the magnitude of
this correction \cite{bijnens93a,baur95a}, see
\cite{leutwyler01a,leutwyler96a} for a discussion. Recent results
from the lattice approach \cite{namekawa05a,yamada05a} seem to indicate
only a mild breaking of Dashen's theorem.

Note that to LO in $\chi$PT using experimental values of the
$\pi$ and $K$ masses, namely
\begin{eqnarray}
   m_{\pi^+}^{\expt} = 139.6\,\mbox{MeV} &
   m_{\pi^0}^{\expt} = 135.0\,\mbox{MeV} \,,
                                           \nonumber \\
   m_{K^+}^{\expt} = 493.7\,\mbox{MeV}   &
   m_{K^0}^{\expt} = 497.7\,\mbox{MeV}  \,,
\label{ps_expt_values}
\end{eqnarray}
we have the result
\begin{eqnarray}
   {m^{\cal S}_s(M) \over m_{ud}^{\cal S}(M)}
      &=& { m_{K^+}^2 + m_{K^0}^2 - m_{\pi^+}^2 \over m_{\pi^+}^2 }
                                           \nonumber \\
      &=& { (m_{K^+}^{\expt})^2 + (m_{K^0}^{\expt})^2 - (m_{\pi^+}^{\expt})^2
                    \over (m_{\pi^0}^{\expt})^2 } \approx 25.9 \,,
\label{LO_standard_ratio}
\end{eqnarray}
independent of the value of $c_a$. So if we are in or close to this regime,
once we have determined the strange quark mass, this immediately gives
an estimate of the light quark mass. Incorporating the NLO terms needs
a determination of all the $c_a$ and $c_i$ ($i=b,c,d$) coefficients
in eqs.~(\ref{strange_general}), (\ref{ud_general}) and gives
the results in section~\ref{q_masses}.


\section{The Lattice Approach}
\label{lattice}


Here we shall derive results for the unquenched ($n_f=2$) strange quark
mass using the axial Ward identity. All our numerical computations
are done with degenerate valence quark masses.

The starting point is the AWI; in the continuum we have
the renormalised relation
\begin{equation}
   \partial_\mu {\cal A}^{\cal R}_\mu
         = 2 m_{q}^{\cal S}(M) {\cal P}^{\cal S}(M) \,,
\end{equation}
where ${\cal A}^{\cal R}$ and ${\cal P}^{\cal S}$ are the renormalised
(in scheme $\cal S$) axial current and pseudoscalar density
respectively. In the (bare) lattice theory the current quark masses
are also defined via the equivalent AWI%
\footnote{$\partial^{\lat}_\mu$ is the symmetric lattice derivative,
conventionally chosen to be
$(\partial^{\lat}_\mu f)(x) = [ f(x+a\hat\mu) - f(x-a\hat\mu) ] /(2a)$,
where $\hat\mu$ is a unit vector in the $\mu$ direction.
\label{footnote_d}}
\begin{equation}
   \partial^{\lat}_\mu {\cal A}_\mu = 2\widetilde{m}_{q} {\cal P} + O(a^2) \,,
\label{pcac_lattice}
\end{equation}
where ${\cal A}$ and ${\cal P}$ are the $O(a)$ improved unrenormalised
axial current and pseudoscalar density
\begin{eqnarray}
   {\cal A}_\mu &=& ( 1 + b_Aam_q ) ( A_\mu + c_A a\partial^{\lat}_\mu P ) \,,
                                           \nonumber \\
   {\cal P}     &=& ( 1 + b_Pam_q ) P \,,
\label{imp_op}
\end{eqnarray}
with
\begin{equation}
   A_\mu = \overline{q}\gamma_\mu\gamma_5 q \,, \qquad 
   P     = \overline{q}\gamma_5 q \,.
\end{equation}
Wilson-type fermions allow several different definitions
of the quark mass. We denote the (bare) quark mass defined from the AWI
with a tilde, $a\widetilde{m}_q$, while that from the VWI is
given by
\begin{equation}
   am_q = {1 \over 2} \left( {1 \over \kappa_q} - {1 \over \kappa^S_{qc}}
                      \right) \,,
\label{vwi_qm}
\end{equation}
where $\kappa_q$ is the Wilson hopping parameter, defining the quark
mass (both sea and valence). The critical sea quark hopping parameter,
$\kappa^S_{qc}$, is defined for fixed $\beta \equiv 6/g_0^2$
(where $g_0$ is the lattice coupling) by the vanishing of the pseudoscalar
mass%
\footnote{We shall suppress the `$V$' superscript on the pseudoscalar mass
and only include an `$S$' superscript where necessary.},
i.e.\ $m_{ps}|_{\kappa_q = \kappa^S_{qc}} = 0$.

Returning to eq.~(\ref{imp_op}), $c_A$ is known non-perturbatively
\cite{dellamorte05b}, but not $b_A$ and $b_P$. However results using
one-loop perturbation theory \cite{sint97a}, or for non-perturbative
quenched QCD \cite{guagnelli00a}, show that the difference $b_A-b_P$
is small (there is however an increase between the perturbative and
quenched non-perturbative results). Multiplying by $am_q$ thus gives
a correction of perhaps half a percent, which with our present level
of accuracy we can ignore.
      
Forming lattice correlation functions means that the quark
mass can be defined and determined from the ratio%
\footnote{Note that to reduce noise, derivatives of operators on the
lattice are taken as compact as possible, consistent with the given
symmetry. Thus we use, no $\mu$ summation, $(\partial_\mu^{2\,\lat} f)(x) =
[ f(x+a\hat\mu) - 2f(x) + f(x-a\hat\mu) ] /(2a)^2
= (\partial_\mu^{\lat}\partial_\mu^{\lat} f)(x) + O(a)$.
\label{footnote_dd}}
\begin{eqnarray}
   a\widetilde{m}_{q}
      &\stackrel{t\gg 0}{=}&
       { \langle
            \partial^{\lat}_4 {\cal A}_4(t){\cal O}(0)
          \rangle \over
            2 \langle {\cal P}(t){\cal O}(0) \rangle }
                                           \nonumber \\
      &\approx&
       { \langle
            \partial^{\lat}_4 A_4(t){\cal O}(0)
          \rangle \over
            2 \langle {\cal P}(t){\cal O}(0) \rangle } +
        c_A a { \langle
                \partial^{2\lat}_4 P(t){\cal O}(0)
                \rangle \over
                2 \langle {\cal P}(t){\cal O}(0) \rangle } + O(a^2)
                                           \nonumber \\
      &\equiv& a\widetilde{m}_q^{(0)} + c_A a\widetilde{m}_q^{(1)} + O(a^2) \,,
\label{correlation_mq}
\end{eqnarray}
where $\approx$ in the second equation signifies that we have dropped
the correction factor $1 + (b_A-b_P)am_q$. ${\cal O}$ is an operator
with a non-zero overlap with the pseudoscalar particle. We choose it
here to be $P^{\smeared}$, where we have used Jacobi smearing
(see the appendix) on the operator.

The parameter space spanned in our numerical simulations is given in
Table~\ref{table_nf2}. The notation is standard for
\begin{table}[t]
   \begin{center}
      \begin{tabular}{||c|l|c|l|l||}
         \hline
         \hline & & &  & \\[-0.75em]
\multicolumn{1}{||c|}{$\beta$}                                & 
\multicolumn{1}{c|}{$\kappa_q^S$}                             & 
\multicolumn{1}{c|}{$c_{sw}$}                                 & 
\multicolumn{1}{c|}{$V$}                                      & 
\multicolumn{1}{c||}{Group}                                   \\
         \hline
 5.20 & 0.1342    & 2.0171 & $16^3\times 32$  & QCDSF  \\
 5.20 & 0.1350    & 2.0171 & $16^3\times 32$  & UKQCD  \\
 5.20 & 0.1355    & 2.0171 & $16^3\times 32$  & UKQCD  \\
         \hline
 5.25 & 0.1346    & 1.9603 & $16^3\times 32$  & QCDSF  \\
 5.25 & 0.1352    & 1.9603 & $16^3\times 32$  & UKQCD  \\
 5.25 & 0.13575   & 1.9603 & $24^3\times 48$  & QCDSF  \\
         \hline 
 5.29 & 0.1340    & 1.9192 & $16^3\times 32$  & UKQCD  \\
 5.29 & 0.1350    & 1.9192 & $16^3\times 32$  & QCDSF  \\
 5.29 & 0.1355    & 1.9192 & $24^3\times 48$  & QCDSF  \\
 5.29 & 0.1359    & 1.9192 & $24^3\times 48$  & QCDSF \\
         \hline
 5.40 & 0.1350    & 1.8228 & $24^3\times 48$  & QCDSF  \\
 5.40 & 0.1356    & 1.8228 & $24^3\times 48$  & QCDSF  \\
 5.40 & 0.1361    & 1.8228 & $24^3\times 48$  & QCDSF  \\
         \hline
         \hline
      \end{tabular}
   \end{center}
\caption{The $\beta$, $\kappa_q^S$ and $c_{sw}$ values%
         \protect\footnotemark\, and the lattice volume
         $V \equiv N_S^3 \times N_T$. The collaboration that
         generated the configurations is given in the last column.}
\label{table_nf2}
\end{table}
\footnotetext{For the number of trajectories generated for each $\kappa_q^S$,
see for example \cite{gockeler05b}.}
the parameters of the action, see for example \cite{gockeler04b}.
(The critical Wilson hopping parameters, $\kappa^S_{qc}$,
have been determined in \cite{gockeler04a} for each $\beta$.)

In the appendix we list our $\kappa_q$ for each $\kappa_q^S$ together
with the corresponding partially quenched $am_{ps}$ and bare
AWI quark masses $a\widetilde{m}_q^{(0)}$, $a\widetilde{m}_q^{(1)}$
and $a\widetilde{m}_q$.


\section{Renormalisation}
\label{renormalisation}


\subsection{Generalities}


Imposing the AWI on the lattice, eq.~(\ref{pcac_lattice}), up to cut-off
effects means that the axial current as well as the pseudoscalar
density and the quark mass must be renormalised. Thus we have
\begin{equation}
   {\cal A}^{\cal R}_\mu = Z_A {\cal A}_\mu \,,  \qquad
   {\cal P}^{\cal S}(M)  = Z^{\cal S}_P(M) {\cal P} \,,
\end{equation}
giving
\begin{equation}
   m_q^{\cal S}(M) = Z^{\cal S}_{\tilde{m}}(M) \widetilde{m}_q  \,,
       \qquad
           Z^{\cal S}_{\tilde{m}}(M) = {Z_A \over Z^{\cal S}_P(M)} \,,
\end{equation}
or in RGI form
\begin{equation}
   m_q^{\rgi} = Z^{\rgi}_{\tilde{m}} \widetilde{m}_q \,, \qquad
                   Z^{\rgi}_{\tilde{m}} =
                     \Delta Z^{\cal S}_m(M) Z^{\cal S}_{\tilde{m}}(M) \,.
\label{rgi_eqn}
\end{equation}


\subsection{Non Perturbative renormalisation}
\label{NP_renormalisation}


We shall employ here the $\rm{RI}^\prime$-MOM scheme
\cite{martinelli94a}, which is easily transcribed to the lattice.
Our implementation of this method is described in
\cite{gockeler98a}. As discussed in section~\ref{rgi}
to obtain the RGI quark mass, we must determine both
$\Delta Z_m^{\ripmom}$ and $Z_{\tilde{m}}^{\ripmom}$.
The RGI quark mass can then be easily converted back to the
$\overline{\rm MS}$ scheme.


\subsubsection{$\Delta Z_m^{\ripmom}$}


We start with $\Delta Z_m^{\ripmom}$.
To write down the perturbative expansion, a definition of the
coupling constant is required. The anomalous dimension coefficients
have been determined to fourth order in
\cite{martinelli94a,franco98a,chetyrkin99a} by taking the
coupling constant to be $g^{\msbar}$. Thus the anomalous 
dimension function is considered as a function of $g^{\msbar}$.
(Other definitions of the coupling constant are possible, more closely
related to MOM schemes \cite{chetyrkin00a}.) So we write
\begin{equation}
   \gamma_m^{\ripmom}(g^{\msbar}) = 
       d_{m0}(g^{\msbar})^2 + d_{m1}^{\ripmom}(g^{\msbar})^4
                                          \ldots \,,
\end{equation}
with coefficients given by \cite{chetyrkin99a},
\begin{eqnarray}
   d_{m1}^{\ripmom}
        &=& - { 2 \over (4\pi)^4}\left[
                    126 - {52 \over 9}n_f
                                 \right] \,,
                                    \nonumber \\
   d_{m2}^{\ripmom} 
        &=& - { 2 \over (4\pi)^6}\left[
                    {20174 \over 3} - {3344 \over 3}\zeta_3
                  - ( {17588 \over 27} - {128 \over 9}\zeta_3 ) n_f
                  + {856 \over 81} n_f^2
                                 \right] \,,
                                    \nonumber \\
   d_{m3}^{\ripmom}
        &=& - { 2 \over (4\pi)^8}\left[
                    {141825253 \over 324} - {7230017 \over 54}\zeta_3
                         + {6160 \over 3}\zeta_5   \right.
                                    \nonumber \\
        & &  \hspace*{1.00in}    \left.
                  - ({3519059 \over 54} - {298241 \over 27}\zeta_3
                      - {4160 \over 3}\zeta_5 ) n_f  \right.
                                    \nonumber \\
        & &  \hspace*{1.00in}    \left.
                  + ( {611152 \over 243} -{ 5984 \over 27}\zeta_3 ) n_f^2
                  - {16024 \over 729} n_f^3 
                                                   \right] \,,
\label{d1+2+3_ripmom}
\end{eqnarray}
which allows $[\Delta Z_m^{\ripmom}(\mu_p)]^{-1}$, where $\mu_p$ is taken
to be the momentum scale in this scheme,  to be computed in the usual way,
\begin{equation}
   [\Delta Z_m^{\ripmom}(\mu_p)]^{-1} = 
          \left[ 2b_0 g^{\msbar}(\mu_p)^2 \right]^{- {d_{m0}\over 2b_0}}
          \exp{\left\{ \int_0^{g^{\msbar}(\mu_p)} d\xi
          \left[ {\gamma_m^{\ripmom}(\xi)
                             \over \beta^{\msbar}(\xi)} +
                 {d_{m0}\over b_0 \xi} \right] \right\} } \,.
\end{equation}
(This result may also be shown by changing the integration variable
in eq.~(\ref{deltam_def}) from some defined
$g^{\ripmom}$ to $g^{\msbar}$, by using eq.~(\ref{G_def}),
i.e.\ $g^{\ripmom}(\mu_p) = G(g^{\msbar}(\mu_p))$.)

\begin{figure}[t]
   \hspace{0.50in}
      \epsfxsize=10.00cm \epsfbox{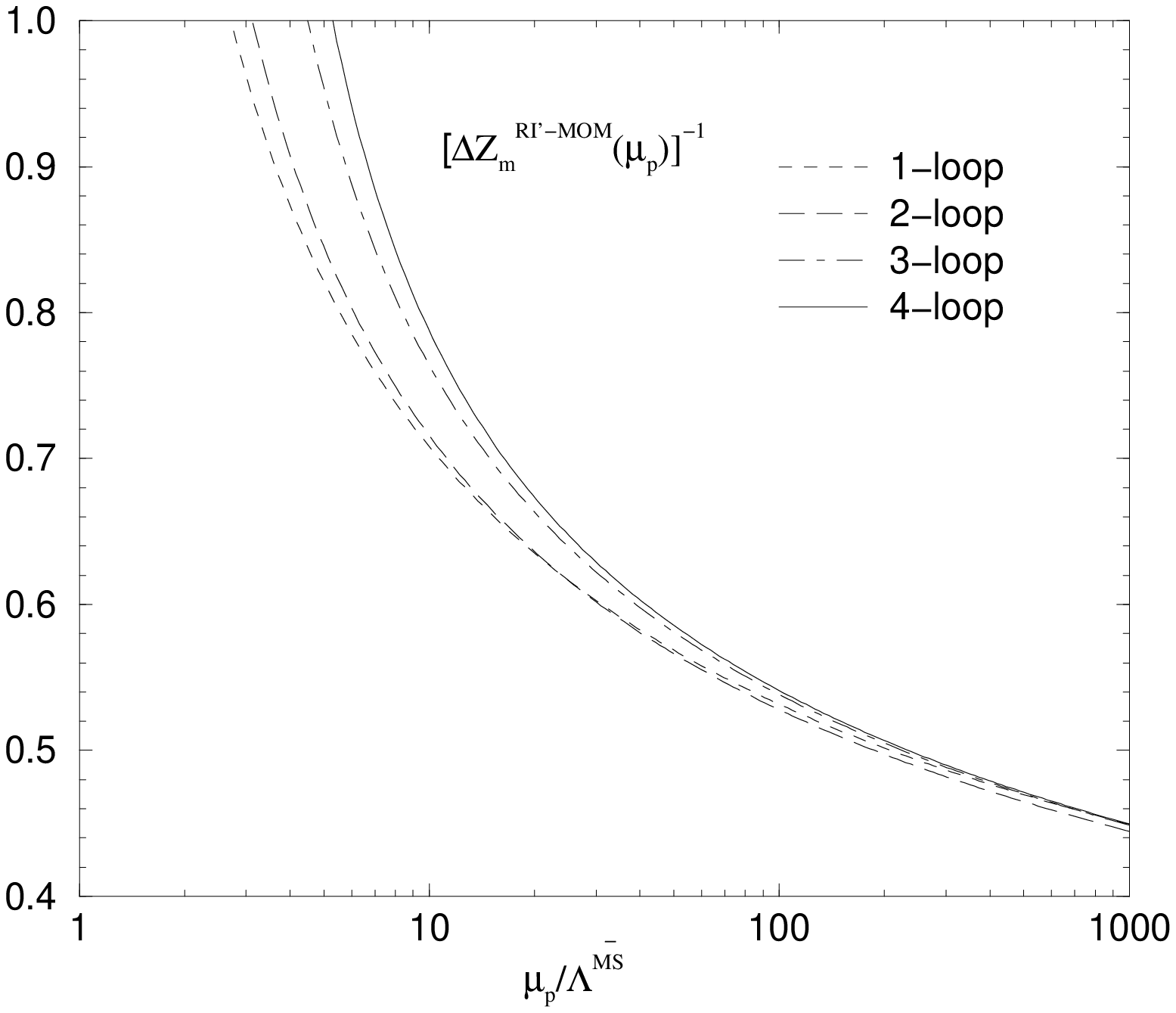}
   \caption{$[\Delta Z_m^{\ripmom}(\mu_p)]^{-1}$ versus
            $\mu_p/\Lambda^{\msbar}$.}
   \label{fig_Del_Zm_MOMms_nf2_muolam}
\end{figure}

In Fig.~\ref{fig_Del_Zm_MOMms_nf2_muolam} we show
$[\Delta Z_m^{\ripmom}]^{-1}$ as a function of $\mu_p / \Lambda^{\msbar}$.
The convergence for $\Delta Z_m^{\ripmom}$ seems slightly worse
in the region of interest than that for the corresponding
$\Delta Z_m^{\msbar}$ as there is more of a change from the 2-loop to
3-loop result (see Fig.~\ref{fig_Del_Zm_MSbar_nf2_muolam} for a comparison
with $\Delta Z_m^{\msbar}$). However this is mitigated by the 
extraction of $\rm{RI}^\prime$-MOM being performed at a range of scales
including higher scales than $2 \, \mbox{GeV}$. Thus, for example,
a typical value of $(a\mu_p)^2 \approx 4$
(see Fig.~\ref{fig_Zm_rgi_b5p20+b5p25+b5p29+b5p40_ap2_050406})
corresponds to $\mu_p / \Lambda^{\msbar} \gsim 20$ where the convergence
(between the 3-loop and 4-loop results) appears to be better.


\subsubsection{$Z_{\tilde{m}}^{\ripmom}$}


The $\rm{RI}^\prime$-MOM scheme considers amputated Green's functions
(practically in the Landau gauge) with an appropriate operator insertion,
here either $A$ or $P$. The renormalisation point is fixed at some
momentum scale $p^2 = \mu_p^2$, and thus we have,
e.g.\ \cite{martinelli94a,gockeler98a},
\begin{eqnarray}
   Z_O^{\ripmom}(\mu_p) 
           = \left. { Z_q^{\ripmom}(p) \over
                        \twelfth \mbox{tr}
                           \left[\Gamma_O(p)
                                 \Gamma_{O,\born}^{-1}(p)
                           \right] } \right|_{p^2=\mu_p^2}
\end{eqnarray}
where $\Gamma_O$ are one-particle irreducible (1PI) vertex functions,
and $Z_q$ is the wave-function renormalisation. We have
generated $\Gamma_O$ only at values of the sea quark mass, i.e.\ the
values in Table~\ref{table_nf2}.

To obtain $Z_{\tilde{m}}^{\ripmom}$ we need both $Z_A$ and $Z_P^{\ripmom}$
in the chiral limit. $Z_A$ is unproblematical, we make a linear fit of
the form
\begin{equation}
   Z_A = A_A + B_A am_q \,,
\label{zachiral_extrp}
\end{equation}
where $am_q$ is defined in eq.~(\ref{vwi_qm}) and $\kappa^S_{qc}$
has been determined in \cite{gockeler04a}.

From the Ward identity obeyed by $\Gamma_P$, we expect, due to chiral
symmetry breaking, that $Z_P^{\ripmom}$ develops a pole in the
quark mass. Hence we have  $Z_P \to 0$ with the quark mass. Thus
following \cite{cudell98a} we try a fit of the form%
\footnote{It made little difference to the $A_P$ (or $A_A)$ coefficients
whether $a\widetilde{m}_q$ or $(am_{ps})^2$ is used.}
\begin{equation}
   (Z_P^{\ripmom})^{-1} = A_P + {B_P \over am_q} \,.
\label{zpchiral_extrp}
\end{equation}
From the operator product expansion \cite{lane74a,pagels79a}
we expect
\begin{equation}
   B_P(\mu_p) \propto { 1 \over (a\mu_p)^2 } \,,
\label{bpope}
\end{equation}
where the constant of proportionality is proportional
to the chiral condensate. Thus we see that
as the scale increases, the $B_P$ coefficient decreases.

A chiral extrapolation using eqs.~(\ref{zachiral_extrp}) and
(\ref{zpchiral_extrp}) has been made using only the sea data sets
to determine the functions $A_A(\mu_p)$ and $A_P(\mu_p)$ respectively.
The lattice momenta originally chosen for the heaviest quark masses
were kept fixed for the extrapolation over the different masses.
If another data set did not have a particular momentum
a linear interpolation was performed between the adjacent momenta
straddling the given momentum.

\begin{figure}[t]
   \hspace{0.50in}
      \epsfxsize=10.00cm 
         \epsfbox{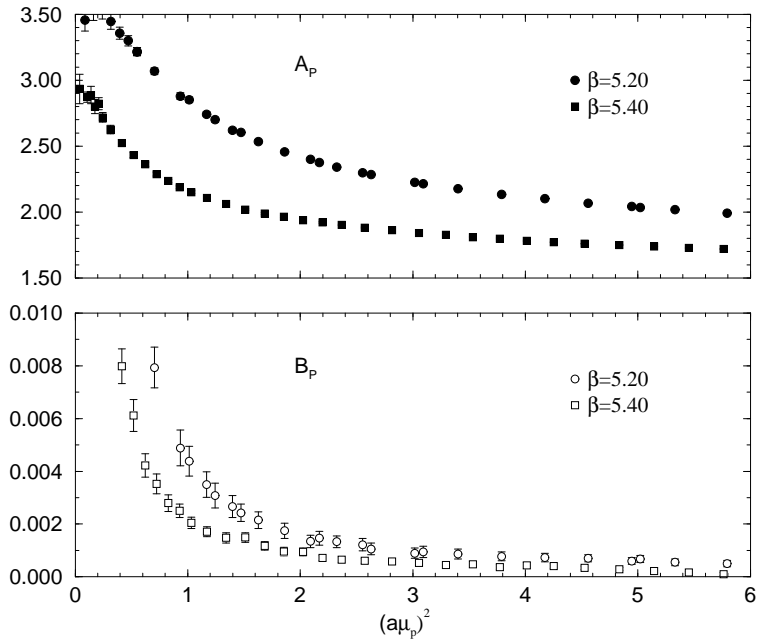}
   \caption{$A_P$ and $B_P$ for $Z_P^{\ripmom}$ for $\beta = 5.20$,
            and $\beta = 5.40$.}
   \label{fig_ooZp_dyn_b5p20+b5p40_amq_chex_ap2_050422}
\end{figure}

Results for $Z_P^{\ripmom}$ are shown for $\beta = 5.20$, $5.40$ in 
Fig.~\ref{fig_ooZp_dyn_b5p20+b5p40_amq_chex_ap2_050422}.
$B_P$ is small and decreasing with increasing $(a\mu_p)^2$
as required from eq.~(\ref{bpope}). (Note that numerically, there is
evidence for a term of the form $B_P/am_q$ as attempting a linear
extrapolation, as for $Z_A$ in eq.~(\ref{zachiral_extrp}),
gave a substantial increase in the fit $\chi^2$.)
Although we do not determine the chiral condensate \cite{becirevic04a}
in this section (see section \ref{generalities}), we note that
numerically the Goldstone pion contamination to $\Gamma_P$ appears
to be small \cite{giusti00a}.


\subsubsection{$Z_{\tilde{m}}^{\rgi}$}
\label{Ztilde_rgi}

\begin{figure}[t]
   \hspace{0.50in} \epsfxsize=9.50cm
      \epsfbox{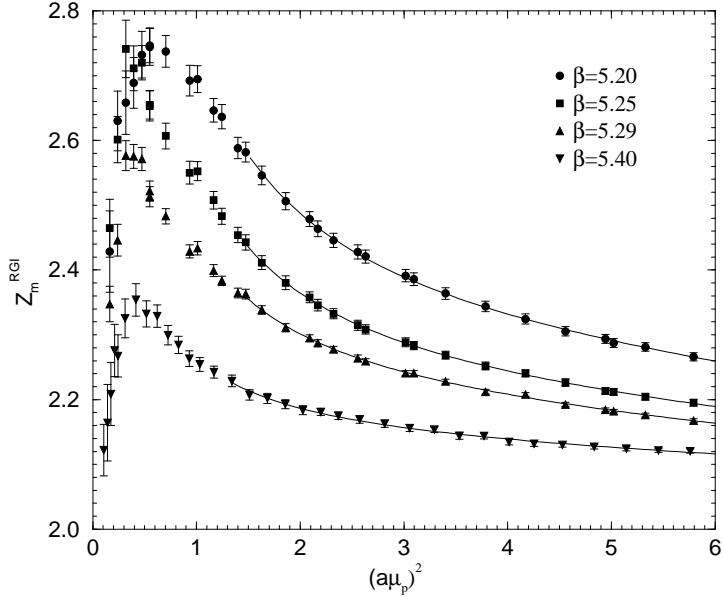}
   \caption{$Z_{\tilde{m}}^{\rgi}$ for $\beta = 5.20$ (filled circles),
            $\beta = 5.25$ (filled squares), $\beta = 5.29$ (filled
            upper triangles), $\beta = 5.40$ (filled lower triangles)
            together with fits as in eq.~(\ref{Zmfit}).}
   \label{fig_Zm_rgi_b5p20+b5p25+b5p29+b5p40_ap2_050406}
\end{figure}

Taking these values for $A_A \equiv Z_A$ and $A_P \equiv Z^{\ripmom}_P$,
forming the ratio $Z_A/Z^{\ripmom}_P$ and multiplying by
$\Delta Z_m^{\ripmom}$ will then give $Z_{\tilde{m}}^{\rgi}$.
This should be independent of $(a\mu_p)^2$. 
Some results are shown in
Fig.~\ref{fig_Zm_rgi_b5p20+b5p25+b5p29+b5p40_ap2_050406}.
Due to cut-off effects, non-perturbative contributions etc.,\
they are not quite constant although the curves become flatter
for increasing $\beta$. To allow for the non-constant remnants
we make a phenomenological fit of the form
\begin{equation}
   F(a\mu_p) = r_1 + r_2 (a\mu_p)^2 + { r_3 \over (a\mu_p)^2 } \,,
\label{Zmfit}
\end{equation} 
where we associate $Z_{\tilde{m}}^{\rgi}$ with $r_1$. This gives the results
in Table~\ref{table_Zmrgi}.
\begin{table}[t]
    \begin{center}
        \begin{tabular}{||c|c||}
           \hline
           \multicolumn{1}{||c|}{$\beta$}      &
           \multicolumn{1}{c||}{$Z_{\tilde{m}}^{\rgi}$}  \\
           \hline
           5.20  & 2.270(12) \\
           5.25  & 2.191(24) \\
           5.29  & 2.177(14) \\
           5.40  & 2.124(06) \\
           \hline
        \end{tabular}
    \end{center}
\caption{Values of $Z_{\tilde{m}}^{\rgi}$ from the NP method of
         section~\ref{Ztilde_rgi}.}
\label{table_Zmrgi}
\end{table}
We start the fit range at $(a\mu_p)^2 = 1.5$, the fit results
for $r_1$ were found to be insensitive to decreasing the fit range.


\subsection{Comparison of $Z_{\tilde{m}}^{\rgi}$ with other results}
\label{comparison_ti}

As many computations of the strange quark mass have used
tadpole improved perturbation theory together with a boosted
coupling constant for the determination of the renormalisation constant,
it is of interest to compare our results obtained in the previous
section with this approach. Our variation of this method,
tadpole-improved renormalisation-group-improved boosted perturbation theory
or TRB-PT, is described in \cite{capitani01a,gockeler04b}. Here
we recapitulate the method. Regarding the lattice as a
`scheme', then from eq.~(\ref{mrgi_msbar}) we can write
\begin{equation}
   m_q^{\rgi} = \Delta Z_{\tilde{m}}^{\lat}(a) \tilde{m}_q(a) \,,
\end{equation}
where the renormalisation-group-improved $\Delta Z_{\tilde{m}}^{\lat}(a)$ 
is given  by eq.~(\ref{deltam_def}). Furthermore in this `lattice' scheme,
we choose to use $g_{\plaq}^2 = g_0^2/u_{0c}^4$ where
$u_0^4 = \langle \third \mbox{Tr}U^{\plaq}\rangle$
($U^{\plaq}$ being the product of links around an elementary plaquette)
rather than $g_0$, as series expansions in $g_{\plaq}$ are believed
to have better convergence. This is boosted perturbation theory.
(We shall use chirally extrapolated plaquette values as determined
in \cite{gockeler05a} at our $\beta$ values and so we add a subscript
`$c$' to $u_0$.) In the tadpole-improved, or mean field approximation,
renormalisation constants for operators with no derivatives are
$\propto u_{0c}$, which indicates that $Z^{\rgi}_{\tilde{m}}u_{0c}^{-1}$
will converge faster than $Z^{\rgi}_{\tilde{m}}$ alone so we re-write
the two-loop equation eq.~(\ref{DelZ_2loop}) as%
\footnote{The TRB-PT subscript in brackets is there only to distinguish
the results from those obtained in section~\ref{NP_renormalisation}.}
\begin{equation}
   Z^{\rgi(\trbpt)}_{\tilde{m}} 
      \equiv \Delta Z^{\lat}_{\tilde{m}}(a)
       = u_{0c}
            \left[ 2 b_0 g^2_{\plaq} \right]^{d_{m0}\over 2b_0}
            \left[ 1 + {b_1 \over b_0} g^2_{\plaq}
            \right]^{{b_0 d_{\tilde{m}1}^{\lat} - b_1 d_{m0}
                                               \over 2 b_0 b_1} +
               { p_1 \over 4 } { b_0 \over b_1} } \,,
\label{DelZ_lat_TI}
\end{equation}
where $p_1$ is the first coefficient in the expansion of $u_{0c}$,
i.e.\ $u_{0c} = 1 - \quart g_0^2 p_1 + \ldots$ with $p_1 = \third$.

It remains to determine $d_{\tilde{m}1}^{\lat}$. This may be found by
relating the (known) perturbative result for $Z_{\tilde{m}}^{\msbar}$ to
$\Delta Z^{\lat}_{\tilde{m}}$ (for simplicity at the scale $\mu = 1/a$) by
\begin{equation}
   Z_{\tilde{m}}^{\msbar}(1/a)
      = { \Delta Z^{\lat}_{\tilde{m}}(a) \over \Delta Z^{\msbar}_m(1/a) }
      = 1 - g_0^2 B^{\msbar}_{\tilde{m}}(1) + \ldots \,,
\label{ZmMSbar_mtwid}
\end{equation}
where
 \begin{equation}
   B^{\msbar}_{\tilde{m}}(c_{sw})
          = { 4/3 \over (4\pi)^2 }
             \left( - 6.79916 + 2.4967c_{sw} - 4.28739c_{sw}^2 \right) \,.
\end{equation}
This result for $B^{\msbar}_{\tilde{m}} \equiv B_A - B_P^{\msbar}$ is
taken from \cite{capitani00a}. (Indeed we could alternatively consider
TRB-PT for $Z_A$ and $Z^{\msbar}_P$ separately and then form the ratio.
The results turned out to be about $1\%$ lower than those presented
here.)

Expanding the ratio in eq.~(\ref{ZmMSbar_mtwid}),
by using the results in eqs.~(\ref{DelZ_lat_TI}) and
(\ref{DelZ_2loop}) for the `lattice' and $\overline{\rm MS}$ schemes
respectively and $t_1$, defined in eq.~(\ref{G_def}) (which using the
notation of \cite{gockeler05a} is numerically given by
$t_1 = t_1^{\lat}(1)$, with
$t^{\lat}_1(c_{sw})  = 0.4682013  - n_f( 0.0066960 - 0.0050467c_{sw} + 
0.0298435c_{sw}^2)$) gives finally
\begin{equation}
   d_{\tilde{m}1}^{\trbpt} = d_{m1}^{\msbar} + d_{m0}(t_1 - p_1) 
                   - 2b_0 B^{\msbar}_{\tilde{m}}(1)
                         \equiv - {4.04873 \over (4\pi)^4} \,.
\label{d1plaq_def}
\end{equation}
Thus from eq.~(\ref{DelZ_lat_TI}) various values of
$\Delta Z^{\lat}_{\tilde{m}}$, or equivalently
$Z^{\rgi(\trbpt)}_{\tilde{m}}$, can be found. Results are given
in Table~\ref{table_Zmrgi_trbpt}.
\begin{table}[t]
    \begin{center}
        \begin{tabular}{||c|c||}
           \hline
           \multicolumn{1}{||c|}{$\beta$}      &
           \multicolumn{1}{c||}{$Z_{\tilde{m}}^{\rgi(\trbpt)}$} \\
           \hline
           5.20  & 1.837  \\
           5.25  & 1.851  \\
           5.29  & 1.862  \\
           5.40  & 1.891  \\
           \hline
        \end{tabular}
    \end{center}
\caption{Values of $Z_{\tilde{m}}^{\rgi(\trbpt)}$ from
         section~\ref{comparison_ti}.}
\label{table_Zmrgi_trbpt}
\end{table}

\begin{figure}[t]
   \hspace{0.50in}
      \epsfxsize=10.00cm 
         \epsfbox{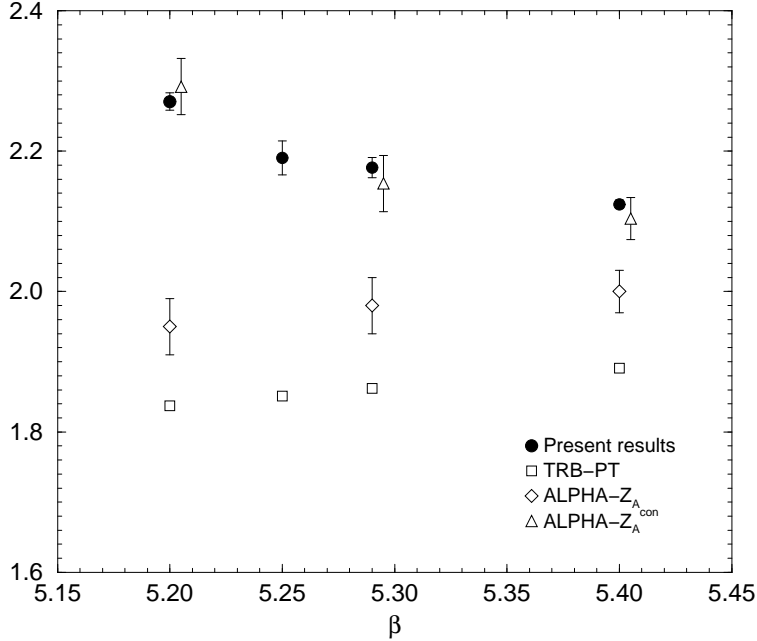}
   \caption{$Z^{\rgi}_{\tilde{m}}$ versus $\beta$. The black circles
            are the results from Table~\ref{table_Zmrgi}, while
            the open squares are the TRG-PT results from
            Table~\ref{table_Zmrgi_trbpt}. Furthermore the open
            diamonds and triangles are the NP results from
            \cite{dellamorte05a}, using the two different results
            for the axial renormalisation constant, \cite{dellamorte05c}.
            (The empty triangle results has been slightly displaced
            for clarity.)}
   \label{fig_Zrgi_mtwid_beta_051024}
\end{figure}

We now turn to a comparison of the results.
In Fig.~\ref{fig_Zrgi_mtwid_beta_051024} we plot $Z^{\rgi}_{\tilde{m}}$
versus $\beta$. Our NP results from section~\ref{NP_renormalisation}
are shown as filled circles. They are to be compared with the TRB-PT
results denoted by empty squares. While there is a difference between
the results, it is decreasing for $\beta \to \infty$ and thus may
be primarily due to remnant $O(a^2)$ effects, which disappear
in the continuum limit.
That various determinations of $Z^{\rgi}_{\tilde{m}}$ have different
numerical values can be seen from the results of \cite{dellamorte05a}
(open diamonds and triangles). In these results two different
definitions of the axial renormalisation constant have been used,
\cite{dellamorte05c}. $Z_A$ is computed when dropping certain
disconnected diagrams, while $Z_A^{\con}$ includes them. The difference
between the two definitions is an $O(a^2)$ effect. Using $Z^{\con}_A$
in $Z^{\rgi}_{\tilde{m}}$ leads, perhaps coincidently, to very similar
results to our NP results.

\begin{figure}[t]
   \hspace{0.50in}
      \epsfxsize=10.00cm 
         \epsfbox{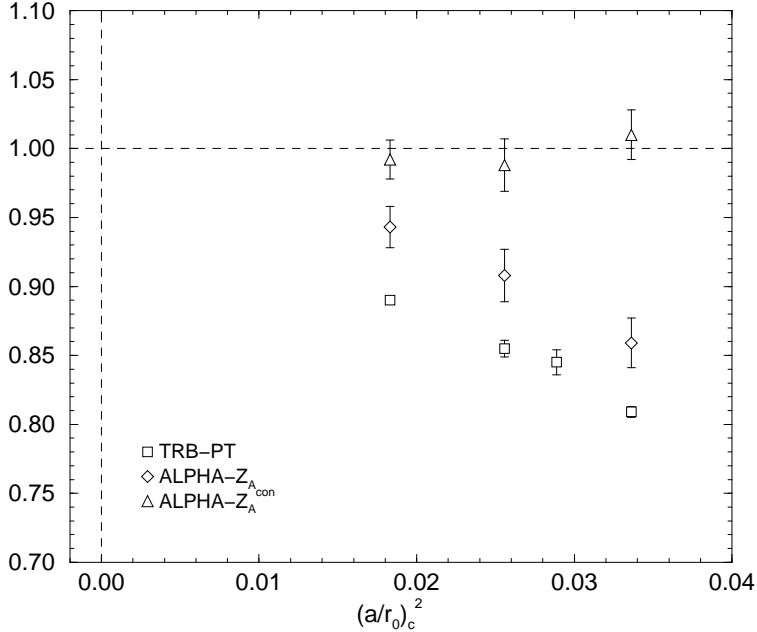}
   \caption{$R_{\tilde{m}}^{X}$ versus $(a/r_0)_c^2$ for $X = \mbox{TRB-PT}$ 
            (open squares), $X = \mbox{ALPHA-$Z_A$}$ (open diamonds)
            and $X = \mbox{ALPHA-$Z_A^{\con}$}$ (open triangles).}
   \label{fig_Zm_rgi_ratio_051026}
\end{figure}

Investigating the possibility of $O(a^2)$ differences a little
further, we note that if we have two definitions of $Z^{\rgi}_{\tilde{m}}$
then if both are equally valid, forming the ratio should yield
\begin{equation}
   R_{\tilde{m}}^X 
     \equiv { Z^{\rgi(X)}_{\tilde{m}} \over  Z^{\rgi}_{\tilde{m}} }
     = 1 + O(a^2) \,,
\end{equation}
where $Z^{\rgi}_{\tilde{m}}$ is the result of
section~\ref{NP_renormalisation} and $X$ is some alternative definition
(i.e.\ TRB-PT, ALPHA-$Z_A$, ALPHA-$Z_A^{\con}$ ).
In Fig.~\ref{fig_Zm_rgi_ratio_051026} we plot this ratio for these
alternative definitions. The $r_0/a$ values used for the $x$-axis are
found by extrapolating the $r_0/a$ results to the chiral limit.
This extrapolation and results for $(r_0/a)_c$ are given in
\cite{gockeler05a}.

We see that (roughly) all three ratios extrapolate to $1$ which implies
that any of the four determinations of $Z^{\rgi}_{\tilde{m}}$ may be
used. This includes the TRB-PT result. Of course other TI determinations
might not have this property, and also their validity always has to be
checked against a NP determination, so this result here is of limited use;
it is always essential to make a NP determination of the renormalisation
constant.
Furthermore it is also to be noted that different determinations can have
rather different $O(a^2)$ corrections, so a continuum extrapolation is 
always necessary.


\section{Results}
\label{results}


\subsection{Generalities}
\label{generalities}

\begin{figure}[p]
   \hspace{0.50in}
      \epsfxsize=10.00cm 
         \epsfbox{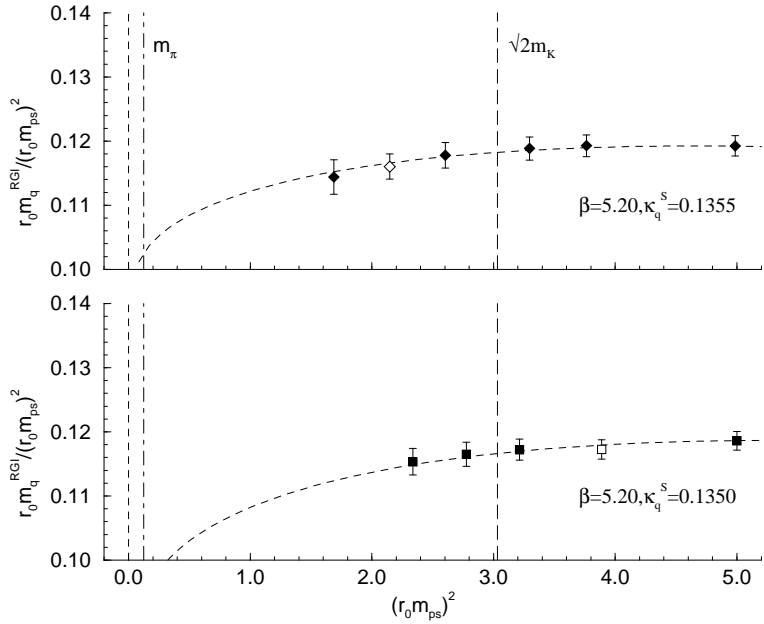}
   \caption{The ratio $r_0m^{\rgi}_q/(r_0m_{ps})^2$
            against $(r_0m_{ps})^2$ for $\beta = 5.20$.
            The fit is described in the text. Results for equal sea
            and valence quark masses are denoted by an open symbol;
            partially quenched results by filled symbols.
            The labelled dashed and dashed-dotted lines are also
            explained in the text.}
   \label{fig_r0mps2_r0mqRGIor0mps2_b5p20_2pic_050921}
\end{figure}
\begin{figure}[p]
   \hspace{0.50in}
      \epsfxsize=10.00cm 
         \epsfbox{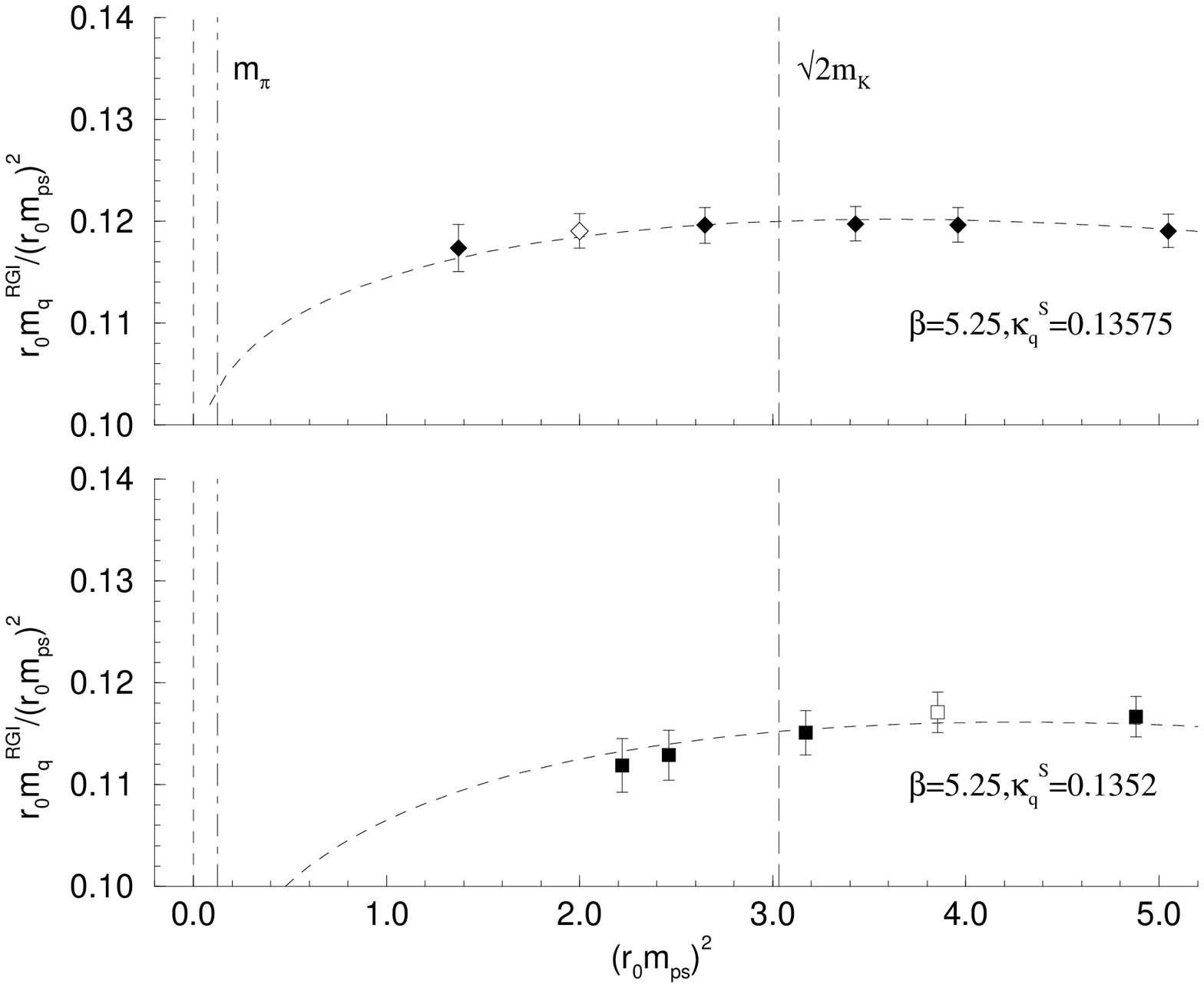}
   \caption{The ratio $r_0m^{\rgi}_q/(r_0m_{ps})^2$
            against $(r_0m_{ps})^2$ for $\beta = 5.25$.
            Notation as for
            Fig.~\ref{fig_r0mps2_r0mqRGIor0mps2_b5p20_2pic_050921}.}
   \label{fig_r0mps2_r0mqRGIor0mps2_b5p25_2pic_050921}
\end{figure}
\begin{figure}[p]
   \hspace{0.50in}
      \epsfxsize=10.00cm 
         \epsfbox{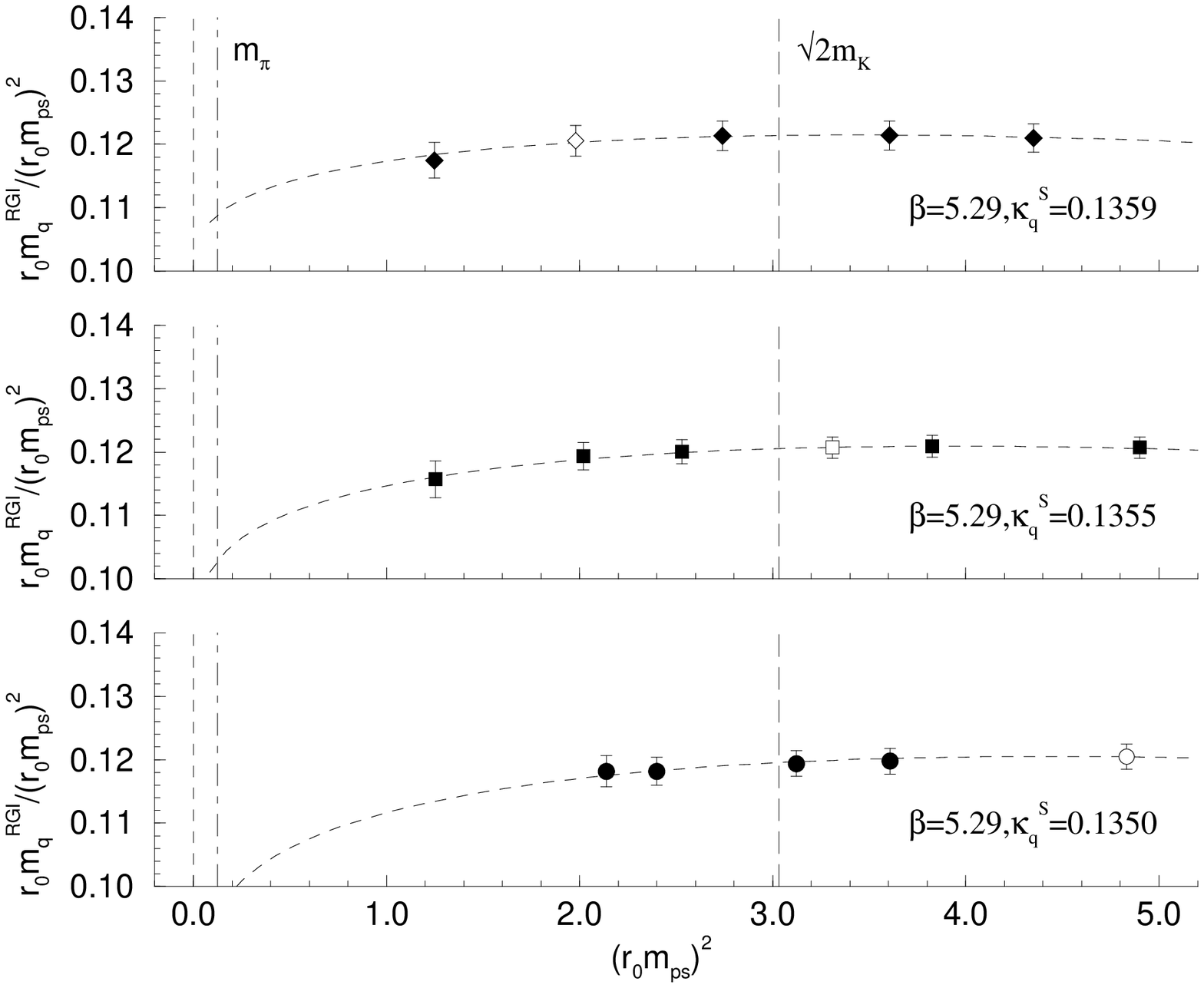}
   \caption{The ratio $r_0m^{\rgi}_q/(r_0m_{ps})^2$
            against $(r_0m_{ps})^2$ for $\beta = 5.29$.
            Notation as for
            Fig.~\ref{fig_r0mps2_r0mqRGIor0mps2_b5p20_2pic_050921}.}
   \label{fig_r0mps2_r0mqRGIor0mps2_b5p29_3pic_051103}
\end{figure}
\begin{figure}[p]
   \hspace{0.50in}
      \epsfxsize=10.00cm
         \epsfbox{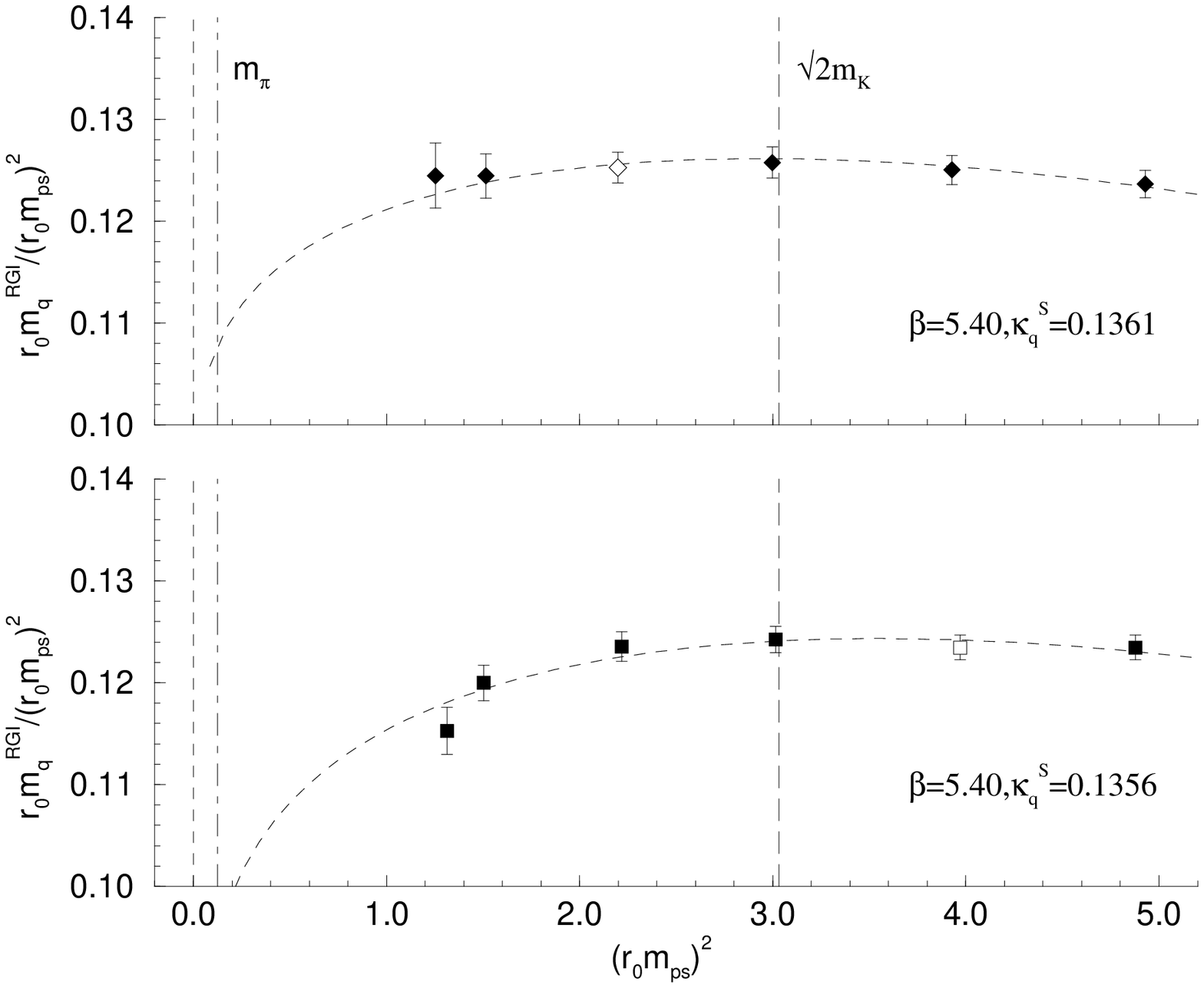}
   \caption{The ratio $r_0m^{\rgi}_q/(r_0m_{ps})^2$
            against $(r_0m_{ps})^2$ for $\beta = 5.40$.
            Notation as for
            Fig.~\ref{fig_r0mps2_r0mqRGIor0mps2_b5p20_2pic_050921}.}
   \label{fig_r0mps2_r0mqRGIor0mps2_b5p40_2pic_050921}
\end{figure}

In Figs.~\ref{fig_r0mps2_r0mqRGIor0mps2_b5p20_2pic_050921},
\ref{fig_r0mps2_r0mqRGIor0mps2_b5p25_2pic_050921},
\ref{fig_r0mps2_r0mqRGIor0mps2_b5p29_3pic_051103} and
\ref{fig_r0mps2_r0mqRGIor0mps2_b5p40_2pic_050921}
we plot the ratio $r_0 m^{\rgi}_q/(r_0m_{ps})^2$ against
$(r_0m_{ps})^2$, where $r_0 \equiv r_0^S$.
(i.e.\ $r_0$ depends only on the sea quark mass, $S$). The numerical
values of $r_0/a$ are given in \cite{gockeler05a}, Table 2.
$r_0$ seems to be a good scale to use because this ratio numerically
does not vary much, as can be seen from the figures.

Using eq.~(\ref{cpt_VV}) (the case of degenerate valence quarks)
we set
\begin{equation}
   y_{V} = r_0 m_q^{\rgi} \,, \quad
   M^S_{ps} = r_0m^S_{ps} \,, \quad
   M^V_{ps} = r_0m_{ps}   \,,
\label{y+M_def}
\end{equation}
to give a fit equation of the form
\begin{eqnarray}
   { r_0 m_q^{\rgi} \over (r_0 m_{ps})^2 }
      = c^{\rgi}_a &+& \hspace*{-0.10in}
        c^{\rgi}_b (r_0 m^S_{ps})^2 +
        c^{\rgi}_c (r_0m_{ps})^2
                                                     \nonumber \\
            &+& \hspace*{-0.10in} 
        c^{\rgi}_d \left( (r_0 m^S_{ps})^2 - 2(r_0 m_{ps})^2 
                               \right) \ln (r_0 m_{ps})^2 \,.
\label{fit_function}
\end{eqnarray}
We use this equation to determine the coefficients $c_a^{\rgi}$
and $c_i^{\rgi}$, $i = b, c, d$. The coefficients depend only
on lattice simulation quantities and the unit chosen, and not on the
scale as  given for example in eq.~(\ref{r0_0.500_phys}) or
eq.~(\ref{r0_0.467_phys}). This can be useful, as an aid, when making
comparisons with other results.

As already mentioned, up to this order in chiral perturbation theory no
input from lattice simulations with non-degenerate valence quark masses
is needed. To test also numerically that the effects from non-degenerate
quark masses are indeed small, we calculated correlation functions
using non-degenerate valence quark masses $m_q^A \ne m_q^B$
as well as degenerate valence quark masses with
$m_q^S \equiv (m_q^A + m_q^B)/2$. We found the relevant quantities
$a m_{ps}$ and $a \widetilde{m}_q$ to differ by $\lsim 1\%$.

In Figs.~\ref{fig_r0mps2_r0mqRGIor0mps2_b5p20_2pic_050921},
\ref{fig_r0mps2_r0mqRGIor0mps2_b5p25_2pic_050921},
\ref{fig_r0mps2_r0mqRGIor0mps2_b5p29_3pic_051103} and
\ref{fig_r0mps2_r0mqRGIor0mps2_b5p40_2pic_050921}
the dashed lines (labelled `$\sqrt{2}m_K$') represent a
fictitious particle composed of two strange quarks, which at LO
$\chi$PT gives from eq.~(\ref{strange_general}) (or equivalently
eq.~(\ref{strange_r0})) the line
$(r_0m_{ps})^2 \equiv (r_0m_{K^+})^2 + (r_0m_{K^0})^2 - (r_0m_{\pi^+})^2$,
while the dashed-dotted lines (labelled `$m_\pi$') represent
a fictitious pion with mass degenerate $u$ and $d$ quarks given
from eq.~(\ref{ud_general}) (or equivalently eq.~(\ref{ud_r0}))
by $(r_0m_{\pi^+})^2$.

The presence of a chiral logarithm in the data manifests
itself in the bending of the results for smaller quark mass,
which can be seen in 
Figs.~\ref{fig_r0mps2_r0mqRGIor0mps2_b5p20_2pic_050921},
\ref{fig_r0mps2_r0mqRGIor0mps2_b5p25_2pic_050921},
\ref{fig_r0mps2_r0mqRGIor0mps2_b5p29_3pic_051103} and
\ref{fig_r0mps2_r0mqRGIor0mps2_b5p40_2pic_050921}.
Results for the fit parameters are given in 
Table~\ref{rgi_fit_results}.
\begin{table}[t]
    \begin{center}
        \begin{tabular}{||c|c|c|c|c||}
           \hline
 $\beta$& $c^{\rgi}_a$&  $c^{\rgi}_b$ &  $c^{\rgi}_c$& $c^{\rgi}_d$  \\
           \hline
   5.20  & 0.1115(53) & -0.00227(187) & 0.00557(491) & 0.00121(133)  \\
   5.25  & 0.1169(42) & -0.00428(160) & 0.00612(399) & 0.00153(111)  \\
   5.29  & 0.1166(29) & -0.00199(117) & 0.00470(287) & 0.00121(090)  \\
   5.40  & 0.1218(24) & -0.00324(079) & 0.00646(179) & 0.00189(056)  \\
           \hline
  $\infty$
         & 0.1330(74) & -0.00378(254) & 0.00789(607) & 0.00275(179)  \\
           \hline
        \end{tabular}
    \end{center}
\caption{Values of $c_a^{\rgi}$ and $c_i^{\rgi}$ ($i = b, c, d$)
         together with their extrapolated (continuum) values
         ($\beta = \infty$).}
\label{rgi_fit_results}
\end{table}

\begin{figure}[t]
   \hspace{0.50in}
      \epsfxsize=10.00cm \epsfbox{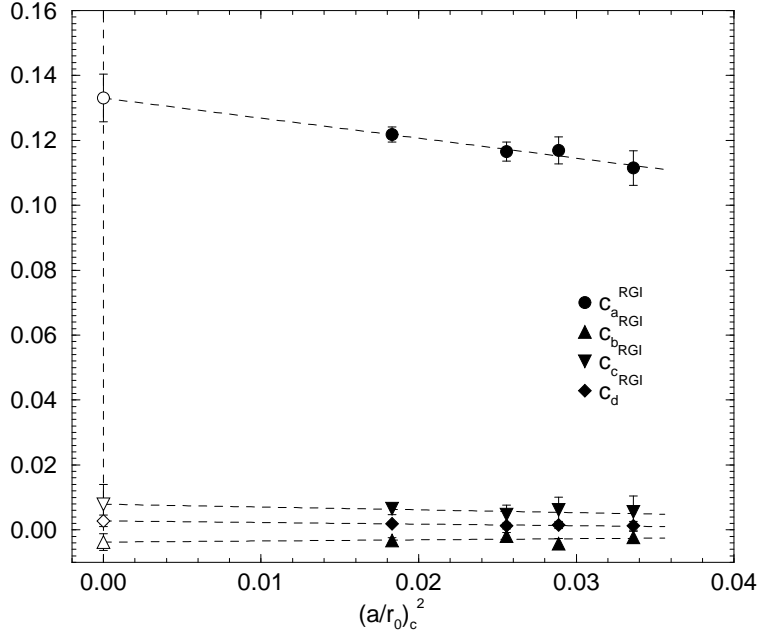}
   \caption{$c_a^{\rgi}$ (filled circles), $c_b^{\rgi}$ (filled upper
            triangles) $c_c^{\rgi}$ (filled lower triangles) and
            $c_d^{\rgi}$ (filled diamonds) versus $(a/r_0)_c^{\:\:2}$.
            The $(r_0/a)_c$ i.e.\ the chirally extrapolated values
            for $r_0/a$ are used. The open symbols represent the
            values of $c_a^{\rgi}$ and $c_i^{\rgi}$ ($i = b, c, d$)
            in the continuum limit.}
   \label{fig_cirgi_051103}
\end{figure}

Numerically we expect the leading order in $\chi$PT to be dominant
with the NLO giving only minor corrections%
\footnote{Alternative plots, using constant $a$ so
$a\widetilde{m}_q/(am_{ps})^2$ against $(am_{ps})^2$ or
equivalently $r_0\widetilde{m}_q/(r_0m_{ps})^2$ against
$(r_0m_{ps})^2$ using $(r_0/a)_c$ which is the chirally extrapolated
$r_0/a$ would give larger NLO corrections.},
i.e.\ we expect $c_a^{\rgi} \gg (r_0m_{ps})^2 c^{\rgi}_i$ ($i = b, c, d$)
and this is indeed found. Using these results for
$c_a^{\rgi}$, $c_i^{\rgi}$ ($i = b, c, d$) in Fig.~\ref{fig_cirgi_051103}
we present their continuum extrapolations.
The $r_0/a$ values used for the $x$-axis are found by extrapolating
the previously used $r_0/a$ results to the chiral limit.
This extrapolation and results (for $(r_0/a)_c$) are also given in 
\cite{gockeler05a}.
This gives values in the last line of Table~\ref{rgi_fit_results}
($\beta = \infty$).

As a first check on these results and to find some idea of possible
systematic effects, we have varied the fit interval from the chosen
$(r_0 m_{ps})^2 \lsim 5$ to $(r_0 m_{ps})^2 \lsim 4$ or $6$ or $\infty$,
i.e.\ include all the data. (We can go no lower than 
$(r_0 m_{ps})^2 \lsim 4$ if we wish to have at least two sea quark masses
in the fit for each $\beta$.) There was little change in
the fit result and we shall take these changes as a systematic error,
see below and section~\ref{q_masses}. Using $(r_0 m_{ps})^2 \lsim 5$
means that our pseudoscalar masses range from about $440\,\mbox{MeV}$
to about double that value.

As a second check on the validity of these results, let us relate
them to the LECs in eqs.~(\ref{LEC_LO_relation}), (\ref{LEC_NLO_cm_relation})
and (\ref{LEC_c_relation}) (in the continuum limit).
From Table~\ref{rgi_fit_results} and eq.~(\ref{LEC_c_relation}) we find that
\begin{equation}
   2\alpha_6 - \alpha_4 \approx -0.21 \,, \quad
   2\alpha_8 - \alpha_5 \approx -1.19 \,.
\end{equation}
These numbers are to be compared with the results of
eq.~(\ref{LEC_alpha}) namely  $\approx -0.24$ and $\approx -1.02$
respectively. Furthermore we have from eqs.~(\ref{chi_def}),
(\ref{chi_AA_def}), (\ref{y_M_defs}) and (\ref{y+M_def}) the relations
\begin{equation}
   c_\chi = {2B_0^{\rgi} \over (4\pi f_0)^2 r_0} \,,\qquad
   c_m    = {1 \over 4\pi f_0 r_0} \,,
\end{equation}
where $B_0^{\rgi} = [\Delta Z_m^{\cal S}(M)]^{-1} B_0^{\cal S}$,
which give
\begin{equation}
   B_0^{\rgi} = { 1 \over 2r_0c_a^{\rgi}} \,.
\end{equation}
Together with eq.~(\ref{chiral_condensate}) (using $f_0 \approx f_\pi$) 
and  $[\Delta Z_m^{\msbar}(2\,\mbox{GeV})]^{-1}$ from
Table~\ref{table_2GeVmsbar_values} this leads to (cf eq.~(\ref{LEC_alpha})),
\begin{equation}
   \langle \overline{q}q \rangle^{\msbar}(2\,\mbox{GeV})
       = - \left\{ \begin{array}{lll}
                      \left(263(5)(5)(5)\,\mbox{MeV}\right)^3 & \mbox{for} &
                          r_0 = 0.5 \,\mbox{fm}   \\
                      \left(267(5)(5)(5)\,\mbox{MeV}\right)^3 & \mbox{for} &
                          r_0 = 0.467 \,\mbox{fm} \\
                   \end{array}
          \right. \,,
\end{equation}
where as discussed in section~\ref{rgi}, we consider two scales
$r_0 = 0.5\,\mbox{fm}$ and $r_0 = 0.467\,\mbox{fm}$. The first error
is statistical and the second is systematic $\approx 5 \,\mbox{MeV}$
determined by the change in $c_a^{\rgi}$ when changing the fit interval,
as discussed above. The third (systematic) error is due to the choice of
$r_0$ scale, taken here as the difference between the results and also
$\approx 5\,\mbox{MeV}$.

We find it encouraging that the $c_a^{\rgi}$ and $c_i^{\rgi}$ ($i = b, c, d$)
results from the fit give numbers in rough agreement with phenomenological
expectations for $2\alpha_6 - \alpha_4$, $2\alpha_8 - \alpha_5$ and
the chiral condensate (and also with other lattice determinations
of the chiral condensate, e.g.\ \cite{becirevic04a,mcneile05a}).
However to obtain more accurate results will require much more
precise numerical data%
\footnote{It is possible to obtain expressions and hence in principle 
results for $f_0$, $\langle \overline{q}q \rangle$,
$2\alpha_6 - \alpha_4$ and $2\alpha_8 - \alpha_5$ in terms of
$c_a^{\rgi}$ and $c_i^{\rgi}$ ($i = b, c, d$). However these then
all depend on the less well determined NLO $c_i^{\rgi}$ ($i = b, c, d$).}.


\subsection{The quark masses}
\label{q_masses}


We now turn to the evaluation of the strange (and light) quark masses.
After finding the coefficients  $c_a^{\rgi}$, $c_i^{\rgi}$ ($i = b, c, d$),
these can be substituted into eq.~(\ref{strange_general}) to give for
the strange quark mass
\begin{eqnarray}
   r_0m_s^{\rgi}
       &=& c^{\rgi}_a 
               \left[ (r_0m_{K^+})^2 + (r_0m_{K^0})^2 - (r_0m_{\pi^+})^2 
               \right]
                                           \nonumber \\
       & & + (c^{\rgi}_b-c^{\rgi}_d)
                \left[(r_0m_{K^+})^2 + (r_0m_{K^0})^2 \right](r_0m_{\pi^+})^2 
                                           \nonumber \\
       & & + \half (c^{\rgi}_c+c^{\rgi}_d)
                \left[(r_0m_{K^+})^2 + (r_0m_{K^0})^2 \right]^2
                                           \nonumber \\
       & & - (c^{\rgi}_b+c^{\rgi}_c)(r_0m_{\pi^+})^4
                                           \nonumber \\
       & & - c^{\rgi}_d \left[(r_0m_{K^+})^2 + (r_0m_{K^0})^2 \right]
                 \left[ (r_0m_{K^+})^2 + (r_0m_{K^0})^2 - (r_0m_{\pi^+})^2
                 \right]
                                           \nonumber \\
       & & \hspace*{1.0in} \times
             \ln \left( (r_0m_{K^+})^2 + (r_0m_{K^0})^2 - (r_0m_{\pi^+})^2
                 \right)
                                           \nonumber \\
       & & + c^{\rgi}_d (r_0m_{\pi^+})^4 \ln (r_0m_{\pi^+})^2 \,.
\label{strange_r0}
\end{eqnarray}
Similarly for the light quark mass we have
\begin{equation}
   r_0m^{\rgi}_{ud} = c_a^{\rgi} (r_0m_{\pi^+})^2 
                      + (c_b^{\rgi}+c_c^{\rgi}) (r_0m_{\pi^+})^4
                      - c_d^{\rgi} (r_0m_{\pi^+})^4 \ln (r_0m_{\pi^+})^2 \,.
\label{ud_r0}
\end{equation}

We first consider the strange quark mass. As can be seen from
Table~\ref{rgi_fit_results} or Fig.~\ref{fig_cirgi_051103},
the errors of the NLO parameters, i.e.\ $c^{\rgi}_i$ ($i = b, c, d$)
are the same size as the signal itself and thus using them directly
in eq.~(\ref{strange_r0}) simply gives a change in the LO result
(i.e.\ using only the $c_a^{\rgi}$ term) of a few percent, together
with a similar increase in the error (especially using error propagation;
note that the third and fifth terms in eq.~(\ref{strange_r0})
give the main contribution to the NLO term).

To reduce the total error on the result, it proved advantageous
to use eq.~(\ref{strange_r0}) to eliminate $c_a^{\rgi}$ from
eq.~(\ref{fit_function}) in terms of
\begin{equation} 
   c^{\rgi}_{a^{\prime}} \equiv
      { r_0m_s^{\rgi} \over
           (r_0m_{K^+})^2 + (r_0m_{K^0})^2 - (r_0m_{\pi^+})^2 } \,.
\end{equation}
This results in a modified fit function of the form
\begin{eqnarray}
   { r_0 m_q^{\rgi} \over (r_0 m_{ps})^2 }
      = c^{\rgi}_{a^{\prime}} &+& \hspace*{-0.10in}
        c^{\rgi}_b [ (r_0 m^S_{ps})^2 - d_b ] +
        c^{\rgi}_c [ (r_0m_{ps})^2 - d_c ]
                                                     \nonumber \\
            &+& \hspace*{-0.10in} 
        c^{\rgi}_d \left[ \left( (r_0 m^S_{ps})^2 - 2(r_0 m_{ps})^2 
                               \right) \ln (r_0 m_{ps})^2 - d_d \right]\,,
\label{modified_fit_function}
\end{eqnarray}
where $d_i$ ($i=b, c, d$) can be read-off from eq.~(\ref{strange_r0})
and have the effect of shifting the various terms in the fit function
by a constant. For example, the simplest to evaluate, $d_b$, is given by
$(r_0m_{\pi^+})^2$. Note that the fit coefficients $c^{\rgi}_i$
($i = b, c, d$) are unchanged from those given in
Table~\ref{rgi_fit_results}. Also the numerical values of the
fit function are unchanged and are given
by the curves in Figs.~\ref{fig_r0mps2_r0mqRGIor0mps2_b5p20_2pic_050921},
\ref{fig_r0mps2_r0mqRGIor0mps2_b5p25_2pic_050921},
\ref{fig_r0mps2_r0mqRGIor0mps2_b5p29_3pic_051103}
and \ref{fig_r0mps2_r0mqRGIor0mps2_b5p40_2pic_050921}.

Note that, although we now have to choose the scale before we
make the fit, the advantage is that the error on $c_{a^{\prime}}^{\rgi}$
gives directly the error on the strange quark mass up to NLO.
Given this, it is no longer a disadvantage to consider the continuum
extrapolation for $m_s^{\msbar}(2\,\mbox{GeV})$ directly.

As discussed earlier, we shall consider two $r_0$ scales.
Given $[\Delta Z_m^{\msbar}(2\,\mbox{GeV})]^{-1}$ from
Table~\ref{table_2GeVmsbar_values}
and using the
experimental values of the $\pi$ and $K$ masses,
eq.~(\ref{ps_expt_values}) to determine the `pure' QCD pseudoscalar
masses in eq.~(\ref{ps_pureQCD_values}), we find the results in
Table~\ref{msbar_fit_results}.
\begin{table}[t]
    \begin{center}
        \begin{tabular}{||c||c|c||l|l||}
           \hline 
  $\beta$& \multicolumn{2}{|c||}{$c^{\rgi}_{a^{\prime}}$}
         & \multicolumn{2}{c||}{$m_s^{\msbar}(2\,\mbox{GeV})$}       \\
           \hline
         & $r_0 = 0.5\,\mbox{fm}$ & $r_0 = 0.467\,\mbox{fm}$ 
         & $r_0 = 0.5\,\mbox{fm}$ & $r_0 = 0.467\,\mbox{fm}$         \\
           \hline
   5.20  & 0.1179(21) & 0.1175(21)  & 98.71(2.33)   & 93.35(2.29)    \\
   5.25  & 0.1233(20) & 0.1231(20)  & 103.27(2.33)  & 97.77(2.30)    \\ 
   5.29  & 0.1216(20) & 0.1214(21)  & 101.79(2.33)  & 96.41(2.31)    \\
   5.40  & 0.1282(18) & 0.1281(18)  & 107.32(2.26)  & 101.73(2.25)   \\
           \hline
  $\infty$
         & 0.1393(46) & 0.1395(46)  & 116.5(5.6)    & 110.7(5.5)     \\
           \hline
        \end{tabular}
    \end{center}
\caption{Values of $c_{a^{\prime}}^{\rgi}$ and
         $m_s^{\msbar}(2\,\mbox{GeV})$ together with their extrapolated
         (continuum) values ($\beta = \infty$).
         $c_i^{\rgi}$ ($i = b, c, d$) are
         given in Table~\ref{rgi_fit_results}.}
\label{msbar_fit_results}
\end{table}

\begin{figure}[t]
   \hspace{0.50in}
      \epsfxsize=10.00cm \epsfbox{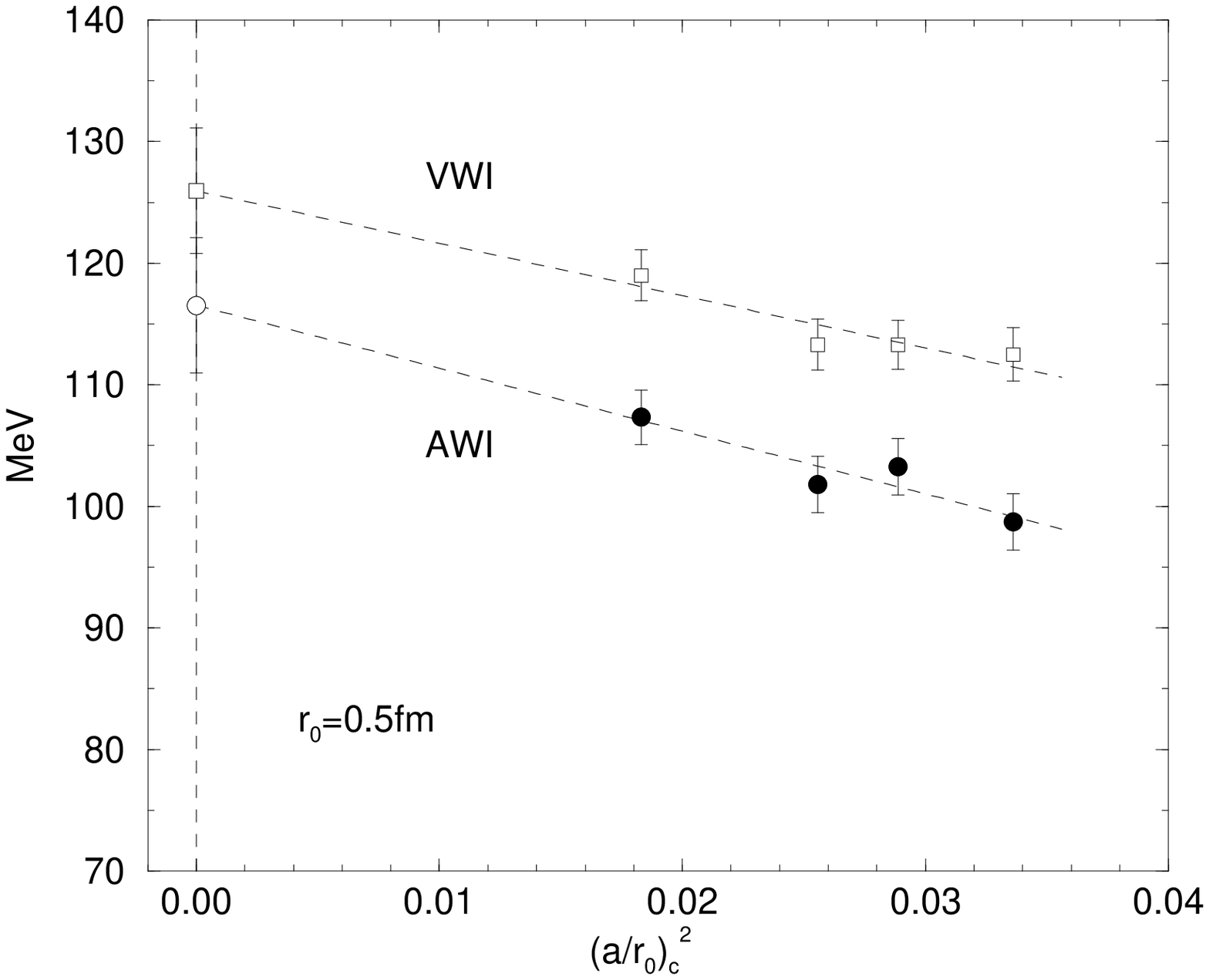}
   \caption{$m_s^{\msbar}(2\,\mbox{GeV})$ versus $(a/r_0)_c^{\:\:2}$
            (filled circles), together with a linear continuum 
            extrapolation (empty circle). The scale used is
            $r_0 = 0.5\,\mbox{fm}$. For comparison we also show the
            results from using the VWI, open squares, \cite{gockeler04a}.}
   \label{fig_msMSbar_AWI+VWI_LO+NLO_051103}
\end{figure}

A continuum extrapolation (using $r_0 = 0.5\,\mbox{fm}$) is shown in
Fig.~\ref{fig_msMSbar_AWI+VWI_LO+NLO_051103} together with a comparison
with our previous VWI results \cite{gockeler04a}. The two methods
have different $O(a^2)$ discretisation errors, but should agree in the
continuum. For the VWI method an extrapolation gave results of
$126(5)(8)\, \mbox{MeV}$, $119(5)(8)\, \mbox{MeV}$
for $r_0 = 0.5\, \mbox{fm}$ and $0.467\, \mbox{fm}$ respectively.
Consistent agreement between the AWI and VWI methods within error bars
is found. Finally we quote our result for the strange quark mass
\begin{equation}
   m_s^{\msbar}(2\,\mbox{GeV})
           = \left\{ \begin{array}{lll}
                        117(6)(4)(6)\,\mbox{MeV} & \mbox{for} &
                          r_0 = 0.5 \,\mbox{fm}   \\
                        111(6)(4)(6)\,\mbox{MeV} & \mbox{for} &
                          r_0 = 0.467 \, \mbox{fm}\\
                       \end{array}
               \right. \,,
\label{ms_MeV_results}
\end{equation}
where the first error is statistical and the second is systematic
$\approx 4\,\mbox{MeV}$. As discussed in section~\ref{generalities},
we have determined it from the effect on $c_{a^\prime}^{\rgi}$ of changing
the fit interval. Furthermore the additional third (systematic) error
due to the $r_0$ scale uncertainty is $\approx 6\,\mbox{MeV}$,
see section~\ref{rgi}.

As discussed earlier, from the values of the
$c^{\rgi}_a$, $c^{\rgi}_i$ ($i = b, c, d$)
coefficients we know that the corrections due to the NLO terms are
small. Using the continuum value of $c^{\rgi}_a$ from
Table~\ref{rgi_fit_results} gives for the LO strange quark mass
$m_s^{\msbar}(2\,\mbox{GeV}) = 111(6)\,\mbox{MeV}$ and
$106(6)\,\mbox{MeV}$ for $r_0 = 0.5\,\mbox{fm}$, $r_0 = 0.467\,\mbox{fm}$
respectively. The difference is about $6 \,\mbox{MeV}$, which
means that NLO terms give about a $5\%$ correction. We have not tried
here to estimate the effects of higher order terms in $\chi$PT,
\cite{bijnens05a}.

For the light quark mass, the numerical situation is more fortunate.
From Table~\ref{rgi_fit_results}, we see that
$|((r_0m_{\pi^+})^2 (c^{\rgi}_b + c^{\rgi}_c))/ c^{\rgi}_a| \approx 0.004$
and similarly for
$|((r_0m_{\pi^+})^2 \ln (r_0m_{\pi^+})^2  c^{\rgi}_d / c^{\rgi}_a)|
\approx 0.005$,
so corrections from LO to NLO $\chi$PT are at the $\half\%$ level
and are negligible here. We shall just quote the LO result of
\begin{equation}
   m_{ud}^{\msbar}(2\,\mbox{GeV})
           = \left\{ \begin{array}{lll}
                        4.30(25)(19)(23)\,\mbox{MeV} & \mbox{for} &
                          r_0 = 0.5 \, \mbox{fm}  \\
                        4.08(23)(19)(23)\,\mbox{MeV} & \mbox{for} &
                          r_0 = 0.467 \, \mbox{fm}\\
                       \end{array}
               \right. \,,
\end{equation}
where again the second error is systematic. The third (systematic)
error is due to the scale $\approx 0.23\,\mbox{MeV}$.

Finally, because the NLO corrections of $\chi$PT are small,
we see that the ratio
\begin{equation}
  {m_s^{\msbar}(2\,\mbox{GeV}) \over m_{ud}^{\msbar}(2\,\mbox{GeV})}
             = 27.2(3.2) \,,
\end{equation}
is close to the LO result, eq.~(\ref{LO_standard_ratio}).


\section{Comparisons and Conclusions}
\label{conclusions}


In this article we have estimated the strange quark mass for two flavour
QCD and found the result in eq.~(\ref{ms_MeV_results}), using
$O(a)$ improved clover fermions and taking into consideration
non-perturbative (NP) renormalisation, the continuum extrapolation
of the lattice results and the use of chiral perturbation theory.
The NLO chiral perturbation theory yields a correction of about $5\%$
to the LO result, and the relevant low energy constants
are in rough agreement with the phenomenological values.

\begin{figure}[t]
   \hspace{0.50in}
      \epsfxsize=13.25cm \epsfbox{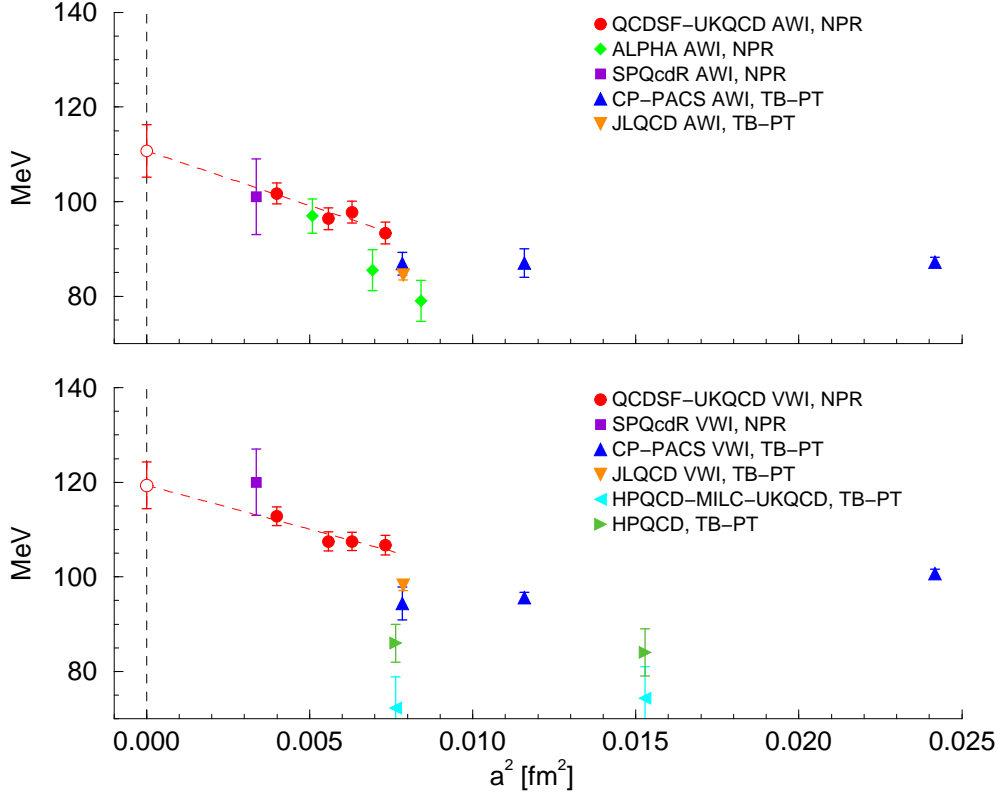}
   \caption{Results for $m_s^{\msbar}(2\,\mbox{GeV})$ versus
            $a^2$ using the AWI (upper plot) and VWI
            (lower plot) methods. The results are presented with the
            collaborations preferred units and scales. Circles
            (together with a linear continuum extrapolation)
            are from this work and \cite{gockeler04a}; diamonds from
            \cite{dellamorte05a}; squares from \cite{becirevic05a};
            up triangles from \cite{alikhan01a}; down triangles from
            \cite{aoki02a}; left triangles from
            \cite{aubin04a}; right triangles from \cite{mason05a}.
            NPR denotes non-perturbative renormalisation, while TB-PT
            denotes TI boosted perturbation theory.
            \cite{aubin04a,mason05a} are for $n_f = 2+1$ flavours;
            the other results are all for $n_f=2$ flavours.}
   \label{fig_comparison_msbar_051116}
\end{figure}

It is also useful to compare our results with the results from other groups.
In Fig.~\ref{fig_comparison_msbar_051116} we show some results for
$n_f=2$ and $n_f=2+1$ flavours (keeping the aspect ratio approximately
the same as in Fig.~\ref{fig_msMSbar_AWI+VWI_LO+NLO_051103}).
A variety of actions, renormalisations, units and scales have been used
(so the results have been plotted in physical units using the authors'
preferred values). In particular the HPQCD-MILC-UKQCD \cite{mason05a}
and HPQCD \cite{aubin04a} collaborations use improved staggered fermions.
These fermions having a (remnant) chiral symmetry and are in the same
situation as overlap/domain wall fermions where there is no distinction
between VWI and AWI quark masses; the bare quark mass in the Lagrangian
simply needs to be renormalised.

As seen earlier in section~\ref{comparison_ti} it is noticeable
that the (tadpole improved) perturbative results lie lower
than the non-perturbatively renormalised results.
Also results with $a \lsim 0.09\,\mbox{fm}$
(i.e.\ $a^2 \lsim 0.008\,\mbox{fm}^2$) appear to be reasonably consistent
with each other (this is more pronounced for the AWI results than for
the VWI results). While results for $a \lsim 0.09$ show some lattice
discretisation effects, using results at larger lattice spacings
seems to give a fairly constant extrapolation to the continuum limit.
A similar effect has also been seen elsewhere, for example in the 
determination of $r_0\Lambda^{\msbar}$ for $n_f=0$ flavours,
\cite{gockeler05a}, where coarse lattices also show this characteristic
flattening of the data.

Finally, we compare these numbers with results from the QCD sum rule
approach. A review of results from this method is given in
\cite{narison05a}, citing as a final result
$m_s^{\msbar}(2\,\mbox{GeV}) = 99(28)\,\mbox{MeV}$, while a recent
five-loop calculation, \cite{chetyrkin05a} gives
$m_s^{\msbar}(2\,\mbox{GeV}) = 105(6)(7)\,\mbox{MeV}$.
These numbers cover the lattice results in
Fig.~\ref{fig_comparison_msbar_051116}.

In conclusion, although there is a spread of results, it would
seem that the unquenched strange quark mass determined here is
not lighter than the quenched strange quark mass and lies in the range of
$100$ -- $130 \, \mbox{MeV}$.


\section*{Acknowledgements}

The numerical calculations have been performed on the Hitachi SR8000 at
LRZ (Munich), on the Cray T3E at EPCC (Edinburgh)
\cite{allton01a}, on the Cray T3E at NIC (J\"ulich) and ZIB (Berlin),
as well as on the APE1000 and Quadrics at DESY (Zeuthen).
We thank all institutions.
This work has been supported in part by
the EU Integrated Infrastructure Initiative Hadron Physics (I3HP) under
contract RII3-CT-2004-506078
and by the DFG under contract FOR 465 (Forschergruppe
Gitter-Hadronen-Ph\"anomenologie).


\clearpage

\appendix

\section*{Appendix}

We collect here in Tables~\ref{table_run_b5p20}, \ref{table_run_b5p25}, 
\ref{table_run_b5p29} and \ref{table_run_b5p40} the numerical values of
the partially quenched $am_{ps}$ and the bare
AWI quark terms $a\widetilde{m}_q^{(0)}$, $a\widetilde{m}_q^{(1)}$
and $a\widetilde{m}_q$. These are defined in eq.~(\ref{correlation_mq})
as ratios of certain correlation functions. (The operators are summed
over spatial planes, the time derivatives being taken as in
footnotes~\ref{footnote_d} and \ref{footnote_dd}.)
The (bootstrap) errors for the ratios are given uniformly
to two significant figures, with the overriding requirement
that the result must also have a minimum of four significant figures.

The second column in the tables gives the pion mass, defined in the
standard way from the correlation function
\begin{equation}
   \langle P^{\smeared}(t) P^{\smeared}(0) \rangle
      \stackrel{t\gg 0}{=} A \left( e^{-m_{ps}t}  +  e^{-m_{ps}(aN_T-t)}
                             \right) \,,
\end{equation}
where the correlation function is evaluated in a configuration
with sea quark mass $am^S_q$. (The normalisation is not important here
so we can work with unimproved operators to obtain the pseudoscalar masses.)
The smearing used is Jacobi smearing (see e.g.\ \cite{best97a}),
with typical parameters $\kappa_s = 0.21$ and $n_s = 50$.

The third and fourth columns in the tables give the bare results for
$a\widetilde{m}_q^{(0)}$ and $a\widetilde{m}_q^{(1)}$, as
defined in eq.~(\ref{correlation_mq}). 

The improvement coefficient $c_A$ has been determined
non-perturbatively in \cite{dellamorte04a}. We use the values
obtained there of
\begin{equation}
   \begin{tabular}{l|l}
      \multicolumn{1}{c}{$\beta$}                             &
      \multicolumn{1}{|c}{$c_A$}                              \\
      \hline
       5.20  &  -0.0641(40) \\
       5.25  &  -0.0565(40) \\
       5.29  &  -0.0517(40) \\
       5.40  &  -0.0420(40) \\
   \end{tabular}
\end{equation}
From the tables it can be seen that the inclusion of the improvement
term ($\times c_A$) to the quark mass gives a noticeable change
in the final result. Also the error in $c_A$ has an effect.
Although not a large difference to using simple error
propagation for the three quantities ($a\widetilde{m}_q^{(0)}$,
$a\widetilde{m}_q^{(1)}$ and $c_A$), to try to minimize the error
propagation we have adopted the procedure of first finding
the bootstrap error for $a\widetilde{m}_q^{(0)} + c_Aa\widetilde{m}_q^{(1)}$
(with fixed $c_A$) and then including the independent error of $c_A$
($\times$ fixed $a\widetilde{m}_q^{(1)}$) by error propagation.
This gives the results for $a\widetilde{m}_q$ shown in the fifth column
of the tables.


\begin{table}[t]
   \begin{center}
      \begin{tabular}{||l|l|l|l|l||}
         \hline
         \hline
\multicolumn{1}{||c}{$\kappa_q$}                              &
\multicolumn{1}{|c}{$am_{ps}$}                                &
\multicolumn{1}{|c}{$a\widetilde{m}^{(0)}_q$}                 &
\multicolumn{1}{|c}{$a\widetilde{m}^{(1)}_q$}                 &
\multicolumn{1}{|c||}{$a\widetilde{m}_q$}                     \\
         \hline
\multicolumn{5}{||c||}{$\kappa_q^S=0.1342$}                   \\
         \hline
     0.1334  &  0.6581(12)  &  0.11221(19) &  0.22501(73) &  0.09778(16) \\
     0.1338  &  0.6224(12)  &  0.10010(19) &  0.20041(70) &  0.08724(16) \\
     0.1342  &  0.5847(12)  &  0.08821(19) &  0.17647(67) &  0.07689(16) \\
     0.1347  &  0.5359(12)  &  0.07366(19) &  0.14744(63) &  0.06420(16) \\
     0.1353  &  0.4720(13)  &  0.05664(19) &  0.11383(60) &  0.04934(16) \\
     0.1356  &  0.4371(14)  &  0.04828(19) &  0.09743(59) &  0.04203(16) \\
     0.1360  &  0.3856(16)  &  0.03715(22) &  0.07576(60) &  0.03229(19) \\
     0.1362  &  0.3569(17)  &  0.03159(25) &  0.06515(64) &  0.02741(22) \\
         \hline
\multicolumn{5}{||c||}{$\kappa_q^S=0.1350$}                   \\
         \hline
     0.1332  &  0.5985(11)  &  0.10076(16) &  0.18529(60) &  0.08888(13) \\
     0.1337  &  0.5515(11)  &  0.08566(16) &  0.15670(58) &  0.07561(13) \\
     0.1342  &  0.5018(11)  &  0.07088(15) &  0.12923(55) &  0.06260(13) \\
     0.1345  &  0.4703(12)  &  0.06218(15) &  0.11327(54) &  0.05492(13) \\
     0.1350  &  0.4148(13)  &  0.04785(15) &  0.08755(54) &  0.04224(13) \\
     0.1353  &  0.3771(15)  &  0.03954(17) &  0.07246(54) &  0.03490(14) \\
     0.1355  &  0.3505(19)  &  0.03397(18) &  0.06256(57) &  0.02996(16) \\
     0.1357  &  0.3216(20)  &  0.02837(20) &  0.05293(57) &  0.02497(17) \\
         \hline
\multicolumn{5}{||c||}{$\kappa_q^S=0.1355$}                   \\
         \hline
     0.1332  &  0.5546(10)  &  0.09121(14) &  0.15890(54) &  0.08102(12) \\
     0.1336  &  0.5158(11)  &  0.07911(14) &  0.13707(52) &  0.07032(13) \\
     0.1340  &  0.4751(11)  &  0.06718(15) &  0.11600(51) &  0.05974(13) \\
     0.1343  &  0.4430(12)  &  0.05842(15) &  0.10069(50) &  0.05196(13) \\
     0.1348  &  0.3848(14)  &  0.04410(15) &  0.07610(51) &  0.03922(14) \\
     0.1350  &  0.3600(15)  &  0.03847(15) &  0.06656(51) &  0.03421(14) \\
     0.1353  &  0.3200(17)  &  0.03014(16) &  0.05252(53) &  0.02677(14) \\
     0.1355  &  0.2907(15)  &  0.02451(15) &  0.04275(43) &  0.02177(13) \\
     0.1357  &  0.2577(23)  &  0.01909(18) &  0.03475(51) &  0.01687(16) \\
         \hline
         \hline
      \end{tabular}
   \end{center}
\caption{The bare results for $am_{ps}$, $a\widetilde{m}_q^{(0)}$,
         $a\widetilde{m}_q^{(1)}$ and $a\widetilde{m}_q$
         for $\beta = 5.20$.}
\label{table_run_b5p20}
\end{table}

\clearpage


\begin{table}[t]
   \begin{center}
      \begin{tabular}{||l|l|l|l|l||}
         \hline
         \hline
\multicolumn{1}{||c}{$\kappa_q$}                              &
\multicolumn{1}{|c}{$am_{ps}$}                                &
\multicolumn{1}{|c}{$a\widetilde{m}^{(0)}_q$}                 &
\multicolumn{1}{|c}{$a\widetilde{m}^{(1)}_q$}                 &
\multicolumn{1}{|c||}{$a\widetilde{m}_q$}                     \\
         \hline
\multicolumn{5}{||c||}{$\kappa_q^S=0.1346$}                   \\
         \hline
     0.1337  &  0.5794(15)  &  0.09697(19) &  0.17225(87) &  0.08723(17) \\
     0.1340  &  0.5514(15)  &  0.08793(19) &  0.15554(86) &  0.07914(17) \\
     0.1346  &  0.4932(10)  &  0.07027(12) &  0.12399(47) &  0.06326(11) \\
     0.1349  &  0.4612(17)  &  0.06148(18) &  0.10782(79) &  0.05539(17) \\
     0.1353  &  0.4168(18)  &  0.05005(18) &  0.08774(76) &  0.04509(17) \\
     0.1355  &  0.3932(18)  &  0.04440(18) &  0.07794(75) &  0.03999(17) \\
     0.1359  &  0.3420(20)  &  0.03317(19) &  0.05877(64) &  0.02985(18) \\
     0.1361  &  0.3133(22)  &  0.02750(21) &  0.04913(65) &  0.02472(19) \\
         \hline
\multicolumn{5}{||c||}{$\kappa_q^S=0.1352$}                   \\
         \hline
     0.1337  &  0.5419(11)  &  0.08860(16) &  0.15186(60) &  0.08002(14) \\
     0.1341  &  0.5027(12)  &  0.07658(16) &  0.13073(57) &  0.06919(14) \\
     0.1345  &  0.4621(13)  &  0.06474(16) &  0.11027(54) &  0.05851(14) \\
     0.1348  &  0.4300(13)  &  0.05599(16) &  0.09536(52) &  0.05060(14) \\
     0.1352  &  0.3821(13)  &  0.04432(12) &  0.07471(38) &  0.04010(10) \\
     0.1355  &  0.3466(17)  &  0.03593(17) &  0.06178(50) &  0.03244(16) \\
     0.1358  &  0.3054(20)  &  0.02740(19) &  0.04775(53) &  0.02470(18) \\
     0.1359  &  0.2901(22)  &  0.02452(21) &  0.04309(56) &  0.02209(20) \\
         \hline
\multicolumn{5}{||c||}{$\kappa_q^S=0.13575$}                  \\
         \hline
     0.1336  &  0.50970(72) &  0.084021(81) &  0.13257(36) &  0.076528(72) \\
     0.1339  &  0.48011(72) &  0.074990(81) &  0.11733(35) &  0.068359(72) \\
     0.1343  &  0.43883(73) &  0.063110(80) &  0.09772(33) &  0.057587(73) \\
     0.1346  &  0.40619(74) &  0.054323(81) &  0.08354(31) &  0.049601(74) \\
     0.1350  &  0.35966(76) &  0.042776(83) &  0.06532(28) &  0.039085(76) \\
     0.1352  &  0.33469(77) &  0.037071(84) &  0.05657(24) &  0.033874(78) \\
     0.1355  &  0.29421(81) &  0.028608(87) &  0.04360(22) &  0.026144(81) \\
     0.13575 &  0.25556(55) &  0.021495(57) &  0.03291(15) &  0.019635(52) \\
     0.1360  &  0.2117(13)  &  0.01456(11)  &  0.02256(26) &  0.013281(98) \\
         \hline
         \hline
      \end{tabular}
   \end{center}
\caption{The bare results for $am_{ps}$, $a\widetilde{m}_q^{(0)}$,
         $a\widetilde{m}_q^{(1)}$ and $a\widetilde{m}_q$
         for $\beta = 5.25$.}
\label{table_run_b5p25}
\end{table}

\clearpage


\begin{table}[t]
   \begin{center}
      \begin{tabular}{||l|l|l|l|l||}
         \hline
         \hline
\multicolumn{1}{||c}{$\kappa_q$}                              &
\multicolumn{1}{|c}{$am_{ps}$}                                &
\multicolumn{1}{|c}{$a\widetilde{m}^{(0)}_q$}                 &
\multicolumn{1}{|c}{$a\widetilde{m}^{(1)}_q$}                 &
\multicolumn{1}{|c||}{$a\widetilde{m}_q$}                     \\
         \hline
\multicolumn{5}{||c||}{$\kappa_q^S=0.1340$}                   \\
         \hline
     0.1340  &  0.5767(11)  &  0.09689(19) &  0.17170(60) &  0.08802(17) \\
     0.1344  &  0.5392(15)  &  0.08480(18) &  0.14841(82) &  0.07713(16) \\
     0.1349  &  0.4901(16)  &  0.07010(18) &  0.12204(81) &  0.06379(16) \\
     0.1352  &  0.4589(17)  &  0.06141(19) &  0.10669(80) &  0.05590(17) \\
     0.1355  &  0.4255(17)  &  0.05283(19) &  0.09169(80) &  0.04809(17) \\
     0.1357  &  0.4024(20)  &  0.04715(19) &  0.08189(80) &  0.04292(17) \\
     0.1359  &  0.3781(21)  &  0.04151(20) &  0.07226(80) &  0.03778(18) \\
     0.1362  &  0.3384(23)  &  0.03315(23) &  0.05830(81) &  0.03013(21) \\
         \hline
\multicolumn{5}{||c||}{$\kappa_q^S=0.1350$}                   \\
         \hline
     0.1340  &  0.52221(81) &  0.08542(11) &  0.14024(41) &  0.078169(94)\\
     0.1343  &  0.49323(83) &  0.07643(11) &  0.12483(40) &  0.069978(96)\\
     0.1347  &  0.45278(86) &  0.06460(11) &  0.10489(39) &  0.059180(99)\\
     0.1350  &  0.42057(92) &  0.05584(11) &  0.09036(34) &  0.051171(95)\\
     0.1355  &  0.3634(10)  &  0.04146(11) &  0.06707(38) &  0.03799(10) \\
     0.1357  &  0.3381(10)  &  0.03577(12) &  0.05794(38) &  0.03277(11) \\
     0.1360  &  0.2963(12)  &  0.02722(12) &  0.04452(39) &  0.02492(11) \\
     0.1361  &  0.2798(17)  &  0.02427(15) &  0.03963(54) &  0.02222(13) \\
         \hline
\multicolumn{5}{||c||}{$\kappa_q^S=0.1355$}                   \\
         \hline
     0.1339  &  0.49968(92) &  0.08260(11) &  0.12790(44) &  0.075988(94) \\
     0.1343  &  0.46105(86) &  0.07062(10) &  0.10825(41) &  0.065025(93) \\
     0.1346  &  0.43015(86) &  0.06175(10) &  0.09402(38) &  0.056888(92) \\
     0.1349  &  0.39774(87) &  0.05297(10) &  0.08023(36) &  0.048825(92) \\
     0.1353  &  0.35144(91) &  0.04142(10) &  0.06252(34) &  0.038192(93) \\
     0.1355  &  0.32688(70) & 0.035783(77) &  0.05417(22) &  0.032983(71) \\
     0.1358  &  0.2858(11)  &  0.02723(11) &  0.04151(30) &  0.02508(10)  \\
     0.1360  &  0.2552(14)  &  0.02159(13) &  0.03316(31) &  0.01988(12)  \\
     0.1363  &  0.2012(18)  &  0.01304(14) &  0.02056(30) &  0.01197(13)  \\
         \hline
\multicolumn{5}{||c||}{$\kappa_q^S=0.1359$}                   \\
         \hline
     0.1339  &  0.47757(65) & 0.078462(92) &  0.11620(29) & 0.072456(88)  \\
     0.13425 &  0.44247(66) & 0.067952(92) &  0.09946(28) & 0.062811(88)  \\
     0.1346  &  0.40540(70) & 0.057566(92) &  0.08333(27) & 0.053259(88)  \\
     0.13505 &  0.35469(70) & 0.044401(93) &  0.06350(25) & 0.041119(89)  \\
     0.13531 &  0.32287(73) & 0.036894(94) &  0.05251(24) & 0.034180(90)  \\
     0.13562 &  0.28151(78) & 0.028036(96) &  0.03982(23) & 0.025977(91)  \\
     0.1359  &  0.23924(87) & 0.020134(92) &  0.02886(21) & 0.018642(86)  \\
     0.13617 &  0.1899(12)  &  0.01239(10) &  0.01837(25) & 0.011444(98)  \\
         \hline
         \hline
      \end{tabular}
   \end{center}
\caption{The bare results for $am_{ps}$, $a\widetilde{m}_q^{(0)}$,
         $a\widetilde{m}_q^{(1)}$ and $a\widetilde{m}_q$
         for $\beta = 5.29$.}
\label{table_run_b5p29}
\end{table}

\clearpage


\begin{table}[t]
   \begin{center}
      \begin{tabular}{||l|l|l|l|l||}
         \hline
         \hline
\multicolumn{1}{||c}{$\kappa_q$}                              &
\multicolumn{1}{|c}{$am_{ps}$}                                &
\multicolumn{1}{|c}{$a\widetilde{m}^{(0)}_q$}                 &
\multicolumn{1}{|c}{$a\widetilde{m}^{(1)}_q$}                 &
\multicolumn{1}{|c||}{$a\widetilde{m}_q$}                     \\
         \hline
\multicolumn{5}{||c||}{$\kappa_q^S=0.1350$}                   \\
         \hline                            
     0.1346  &  0.44399(52) &  0.071849(57) &  0.10023(25) &  0.067635(52) \\
     0.1350  &  0.40301(43) &  0.059913(50) &  0.08250(18) &  0.056444(47) \\
     0.1353  &  0.37156(54) &  0.051132(70) &  0.06996(21) &  0.048190(67) \\
     0.1357  &  0.32541(63) &  0.039519(68) &  0.05359(21) &  0.037266(64) \\
     0.13602 &  0.28482(69) &  0.030274(70) &  0.04100(21) &  0.028550(66) \\
     0.1363  &  0.24504(77) &  0.022220(72) &  0.03027(20) &  0.020947(68) \\
     0.13655 &  0.20349(95) &  0.014975(74) &  0.02098(22) &  0.014093(70) \\
     0.1366  &  0.1934(13)  &  0.013555(84) &  0.01950(39) &  0.012735(79) \\
         \hline
\multicolumn{5}{||c||}{$\kappa_q^S=0.1356$}                   \\
         \hline
     0.1346  &  0.42009(66) &  0.067720(51) &  0.08973(29) &  0.063947(47) \\
     0.13494 &  0.38581(63) &  0.057620(49) &  0.07533(28) &  0.054453(45) \\
     0.1353  &  0.34617(72) &  0.046987(49) &  0.06068(26) &  0.044436(46) \\
     0.1356  &  0.31232(67) &  0.038239(49) &  0.04926(22) &  0.036168(44) \\
     0.13591 &  0.27210(77) &  0.029197(52) &  0.03725(24) &  0.027631(50) \\
     0.13618 &  0.23346(87) &  0.021380(56) &  0.02742(23) &  0.020227(53) \\
     0.13643 &  0.1921(10)  &  0.014081(68) &  0.01873(22) &  0.013293(65) \\
     0.1365  &  0.1796(12)  &  0.01185(14)  &  0.01629(34) &  0.01116(13)  \\
         \hline
\multicolumn{5}{||c||}{$\kappa_q^S=0.1361$}                   \\
         \hline
     0.1346  &  0.40055(60) &  0.064373(49) &  0.08174(23) &  0.060936(45) \\
     0.13493 &  0.36621(63) &  0.054572(47) &  0.06820(22) &  0.051704(44) \\
     0.13525 &  0.33068(70) &  0.045072(53) &  0.05559(21) &  0.042735(50) \\
     0.13555 &  0.29521(76) &  0.036298(51) &  0.04425(20) &  0.034437(49) \\
     0.13584 &  0.25784(85) &  0.027842(55) &  0.03372(20) &  0.026425(53) \\
     0.1361  &  0.22081(72) &  0.020335(47) &  0.02455(16) &  0.019303(44) \\
     0.13632 &  0.1833(12)  &  0.013937(71) &  0.01705(22) &  0.013220(68) \\
     0.1364  &  0.1668(19)  &  0.011552(88) &  0.01457(22) &  0.010940(84) \\
         \hline
         \hline
      \end{tabular}
   \end{center}
\caption{The bare results for $am_{ps}$, $a\widetilde{m}_q^{(0)}$,
         $a\widetilde{m}_q^{(1)}$ and $a\widetilde{m}_q$
         for $\beta = 5.40$.}
\label{table_run_b5p40}
\end{table}

\clearpage



\end{document}